\documentclass[twocolumn, prb, aps, superscriptaddress, longbibliography, floatfix]{revtex4-1}
\pdfoutput=1
\usepackage[caption=false,subrefformat=parens,labelformat=parens]{subfig}
\usepackage{graphicx}
\usepackage{float}
\usepackage{amsmath}
\usepackage{bm}
\usepackage{xcolor}
\usepackage{siunitx}

\usepackage[colorlinks=true, allcolors=blue]{hyperref}
\usepackage{color}

\def\blue{\color{blue}}

\begin{document}

\title{Non-local correlations in Iron Pnictides and Chalcogenides}

\author{Shinibali Bhattacharyya}
\affiliation{Department of Physics, University of Florida, Gainesville, Florida 32611, USA}
\author{Kristofer Bj\"ornson}
\affiliation{Niels Bohr Institute, University of Copenhagen, Lyngbyvej 2, DK-2100 Copenhagen, Denmark}
\author{Karim Zantout}
\affiliation{Institut f\"ur Theoretische Physik, Goethe-Universit\"at, 60438 Frankfurt am Main, Germany}
\author{Daniel Steffensen}
\affiliation{Niels Bohr Institute, University of Copenhagen, Lyngbyvej 2, DK-2100 Copenhagen, Denmark}
\author{Laura Fanfarillo}
\affiliation{Department of Physics, University of Florida, Gainesville, Florida 32611, USA}
\affiliation{Scuola Internazionale Superiore di Studi Avanzati (SISSA), Via Bonomea 265, 34136 Trieste, Italy}
\author{Andreas Kreisel}
\affiliation{
Institut f\" ur Theoretische Physik, Universit\"at Leipzig, D-04103 Leipzig, Germany}
\author{Roser Valent\'\i}
\affiliation{Institut f\"ur Theoretische Physik, Goethe-Universit\"at, 60438 Frankfurt am Main, Germany}
\author{Brian M. Andersen}
\affiliation{Niels Bohr Institute, University of Copenhagen, Lyngbyvej 2, DK-2100 Copenhagen, Denmark}
\author{P. J. Hirschfeld}
\affiliation{Department of Physics, University of Florida, Gainesville, Florida 32611, USA}

\date{\today}
\begin{abstract}
Deviations of low-energy electronic structure of iron-based superconductors from density functional theory predictions have been parametrized  in terms of band- and orbital-dependent mass renormalizations and energy shifts. The former have typically been described in terms of a local self-energy within the framework of dynamical mean field theory, while the latter appears to require non-local effects due to interband scattering. By calculating the renormalized bandstructure in both random phase approximation (RPA) and the two-particle self-consistent approximation (TPSC),
we show that correlations in pnictide systems like LaFeAsO and LiFeAs can be described rather well by a non-local self-energy. In particular, Fermi pocket shrinkage as seen in experiment occurs due to repulsive interband finite-energy scattering. For the canonical iron chalcogenide system FeSe in its bulk tetragonal phase, the situation is however more complex since even including momentum-dependent band renormalizations cannot explain experimental findings. We propose that the nearest-neighbor Coulomb interaction may play an important role
in band-structure renormalization in FeSe. We further compare our evaluations of non-local quasiparticle scattering lifetime within RPA and TPSC with experimental data for LiFeAs.

\end{abstract}


\maketitle
\section{Introduction}

\begin{figure}[hbt!]
    \centering
    \subfloat[]{\includegraphics[width=0.5\linewidth]{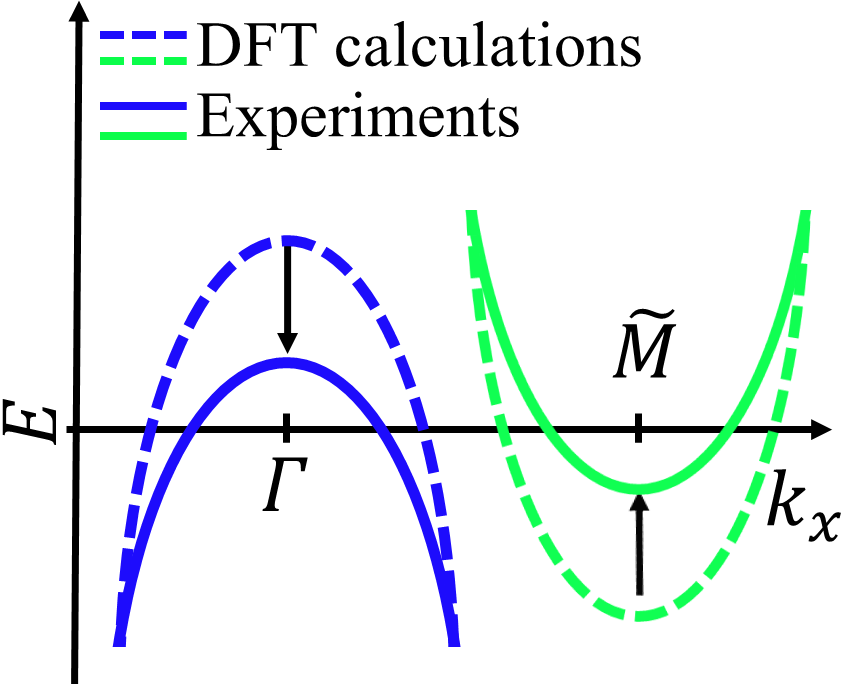}\label{schematic_shrinkage}} \; \;
    \subfloat[]{\includegraphics[width=0.43\linewidth]{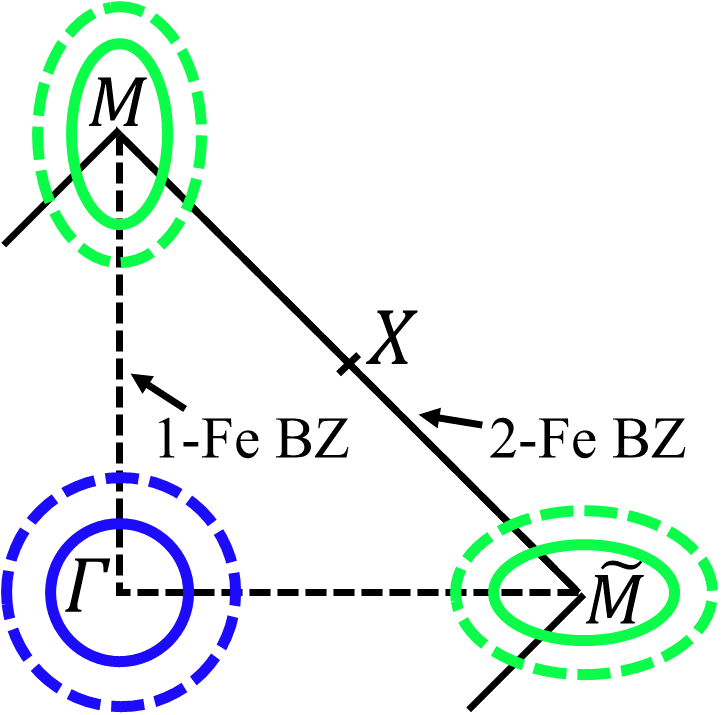}\label{schematic_FS}}
    \caption{Schematic plot showing (a) band structure along $k_x$ direction and (b) corresponding Fermi surface of a generic FeSC in the 1-Fe BZ. The dashed lines represent DFT-predicted $\Gamma$-centered hole (blue) and $\tilde{\textnormal{M}}$-centered electron (green) bands. The solid lines represent the renormalized bands, as observed in dHvA and ARPES experiments, with momentum-dependent energy shifts and shrunken pocket sizes.}
    \label{schematic_cartoons}
\end{figure}

Soon after the discovery of iron-based superconductors (FeSC), it was realized that density functional theory (DFT) calculations of the electronic structure of these materials gave results that were qualitatively in agreement to angle-resolved photoemission spectroscopy (ARPES): the systems consisted of nearly compensated, quasi 2-dimensional hole and electron pockets centered at the high-symmetry points of the Brillouin zone (BZ)~\cite{Lebegue2007,Singh2008,Cao2008, LaFeAsO_Arpes_PRB2010}. At the same time, quantitative discrepancies were noted: Fermi velocities and pocket sizes observed by de Haas-van Alphen (dHvA) and ARPES measurements were smaller as compared to those calculated by DFT~\cite{Coldea2008PRL,
BorisenkoLiFeAs, Lee_etal_Kotliar2012, PutzkePRL2012, Brouet2013PRL,
Coldea2016, Watson2016, Reiss2017}. For the purpose of better visualization, a schematic plot of the band structure and Fermi surface (FS) of a generic FeSC, showing the discrepancy between DFT predictions and experimental findings  is shown in Fig.~\ref{schematic_cartoons}. The ``pocket shrinkage" was explained early on in terms of a rather simple picture.  In a toy multiband model for FeSC, self-energy renormalizations induced by repulsive interband interactions were shown to lead generically to simultaneous shrinkage of both electron and hole pockets~\cite{Ortenzi2008,Fanfarillo2016}, also called ``red-blue shifts"~\cite{Borisenko_FeSePRB2017}. 

While model calculations are valuable, they cannot distinguish easily among different materials; clearly it is desirable to understand effects of correlations within a material-specific first principles method beyond DFT.  Early attempts by dynamical mean field theory (DFT+DMFT) to explain overall materials trends in Fe-based systems with local correlations were quite successful~\cite{Yin2011}, and led to a  picture where inter-orbital Hund's exchange suppressed  inter-orbital charge fluctuations and promoted strong orbital selective correlations, the so-called ``Hund's metal" state \cite{Haule_2009,deMedici_review,Biermann_review}. Subsequent DFT+DMFT
calculations of the FS of LiFeAs, on which high-resolution ARPES data were
available, were also reported as  successful because some pocket shrinkage was
observed~\cite{Ferber2012,Lee_etal_Kotliar2012, Tomczak2012, Tomczak2014}, but on closer examination these calculations did not account for the simultaneous shrinkage of both the
electron and hole pockets. The first full-fledged numerical calculation that discussed band renormalization effects in FeSC was performed in the
framework of the fluctuation exchange approximation (FLEX)~\cite{AritaIkeda2010}. However, the extent of FS shrinkage reported in
such calculations were uncontrollably large which required imposing \textit{ad
hoc} restrictions on the renormalization scheme.

Discrepancies between DFT, DMFT and experimental low-energy band structures were brought into stark relief in the case of FeSe.  DFT calculations
within the local density approximation (LDA) or the generalized approximation
(GGA) show FS with two inner $d_{xz/yz}$ $\Gamma$-centered hole pockets
surrounded by an outer $d_{xy}$ pocket, and two $M$-centered electron pockets
of sizes similar to those found in DFT calculations for other FeSC.  On the
contrary, ARPES at $90$K, just above the tetragonal-to-orthorhombic transition,
finds dramatically smaller pockets, with only a single hole pocket at $\Gamma$
of $d_{xz/yz}$ character~\cite{YiPRX2019}. While it is proposed that the two
$d_{xz/yz}$ bands can split due to spin-orbit coupling~\cite{BorisenkoSO} or
orthorhombic distortions~\cite{wang2019fese}, it is striking that the $d_{xy}$
band appears to be pushed
down by as much as 50 meV compared to
reported DFT calculations. Reported DFT+DMFT calculations seem to also
fail to capture the strong pocket shrinkage and do not reproduce the 
suppression of the $d_{xy}$
band below the Fermi level~\cite{Aichhorn2010,Watson_hubbard2017,Skornyakov2017}. At lower temperatures, the FeSe system becomes
strongly nematic without long range magnetism, an effect that was not found in
any first principles calculation until quite recently~\cite{Long2019} with a
combination of LDA+U and an enlarged unit cell scheme, whose generality is
currently unclear.

Recently, Zantout \textit{et al.}~\cite{ZantoutPRL2019} proposed that a local approximation of correlations was not sufficient to account for band renormalizations in the Fe-pnictides.  They showed that a significantly better description of the relative pocket shrinkage and deformations in LiFeAs band structure was obtained within a multi-orbital formalism of the two-particle self-consistent scheme (TPSC)~\cite{Vilk_Tremblay_TPSC}. TPSC has the advantage that it preserves certain local spin and charge sum rules that are not accounted for in the more popular approximation schemes. The good agreement found with band renormalization and scattering lifetimes from ARPES experiments, left open the question of whether this success was due to some aspect of the particular approximation scheme employed, or to the general inclusion of non-local correlations.  It is therefore important to compare different approximate approaches to calculating non-local self-energy to see the extent to which they agree with experimental evidences, and with each other.

It is also very important to consider and understand the origin of the apparent profoundly different effects of correlations in the two classes of iron pnictides (Pn) and chalcogenides (Ch). In this work, we present calculations of the non-local electronic self-energy within the random phase approximation (RPA) and compare with TPSC calculations. For historical reasons, we also compare with FLEX calculations, but in contrast to  the FLEX approach in Ref.~\onlinecite{AritaIkeda2010}, we ignore first-order Hartree-Fock contributions to the self-energy since our DFT-derived tight-binding parameters already account for important Hartree-Fock corrections. 

This paper is organized as follows: first, we introduce in Section \ref{toy_model} the standard mechanism for Fermi pocket shrinkage in FeSC, as originally discussed by Ortenzi \textit{et al.}~\cite{Ortenzi2008}.  We point out that while their description of the repulsive interband scattering driving the pocket shrinkage is generally correct, the interplay between momentum-transfer and finite-energy scattering processes is somewhat subtle and needs to be considered carefully in the context of realistic models driven by spin-fluctuation interaction. In Section \ref{model}, we introduce the multiorbital Hubbard-Hund Hamiltonian with which realistic  calculations of the self-energy and renormalized band structure will be performed via RPA scheme. In Section \ref{bs_results}, we present RPA results for electronic structure renormalization for LaFeAsO, LiFeAs and  FeSe, which agrees very well with TPSC and FLEX evaluations. We show that in the case of LiFeAs and LaFeAsO (Pn), all the methods agree semi-quantitatively, giving rise to non-local band renormalizations and pocket shrinkage similar to ARPES findings. On the other hand for FeSe (Ch), while results of the three methods are still similar when starting with the same {\it ab-initio} model, 
they remain in dramatic disagreement with the experimental band structure. We show in Section \ref{longrangeCoulomb} that one of the possible physical mechanisms that could explain drastic band renormalization for FeSe is significant nearest-neighbor Coulomb interactions. In Section \ref{comparison}, we explicitly benchmark the results for the static self-energy obtained from RPA, TPSC, and FLEX. We present in Section \ref{scattering}, our results for the quasiparticle lifetimes in LiFeAs at different points on its renormalized FS and compare with ARPES findings.  Finally, we present our conclusions in Section \ref{conclusions}.

\section{Background: Toy model illustrating Fermi surface shrinkage} \label{toy_model}

The phenomenon of Fermi pocket shrinkage, illustrated in Fig.~\ref{schematic_cartoons}, has been discussed in the context of a simple two-band model for FeSC by Ortenzi \textit{et al.} in Ref.~\onlinecite{Ortenzi2008}. They analyzed the changes of the low-energy effective model induced by the coupling to collective modes, described within an Eliashberg framework via a self-energy function for each band. They showed that the multiband character of the electronic structure and the strong particle-hole asymmetry of the bands induce self-energy effects that lead to the shrinking of the FS.

The FS of the interacting system is given by the pole of the renormalized Green's function $ G(\mathbf{k},\omega_{m}) = ( G^0 (\mathbf{k},\omega_{m})^{-1} - \Sigma(\mathbf{k},\omega_m) )^{-1}$, as obtained from Dyson's equation, where $G^0(\mathbf{k},\omega_{m})$ is the non-interacting Green's function and $\Sigma(\mathbf{k},\omega_m)$ the self-energy.  Here $ \omega_{m} = (2m+1)\pi T $ are the fermionic Matsubara frequencies for a given temperature $T$. The renormalized dispersion arises from modification of the DFT dispersion caused by the real part of the static self-energy $\Sigma'(\mathbf{k},\omega_0)$ as $T\rightarrow 0$. A momentum-dependent self-energy that has a negative (positive) value of $\Sigma'(\mathbf{k},\omega_0)$ around the hole (electron) pocket can result in simultaneous pocket shrinkage. Hence, for our purpose, we can investigate the sign of $\Sigma'(\mathbf{k},\omega_0)$. 

\begin{figure}[t!]
     \centering
     \includegraphics[width=\linewidth]{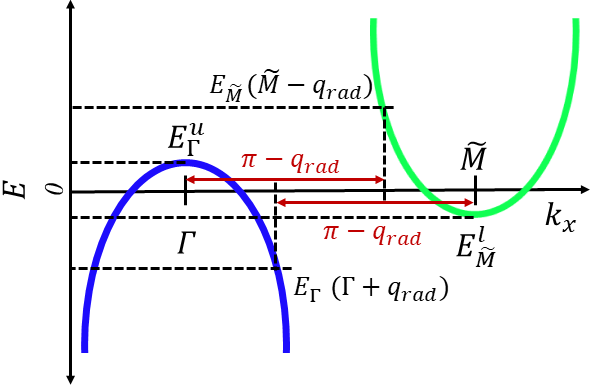}
     \caption{Schematic plot of $\Gamma$-centered hole band and $\tilde{\textnormal{M}}$-centered electron band in 1-Fe zone of a generic FeSC.}
     \label{toy_elhl}
\end{figure}

Revisiting the shrinking mechanism discussed by Ortenzi \textit{et al.}, it is instructive to consider the result of lowest order perturbation theory for a two-band model as shown in Fig.~\ref{toy_elhl}, with a $\Gamma=(0,0)$ centered hole pocket and an $\tilde{\textnormal{M}}=(\pi,0)$ centered electron pocket. We assume purely two-dimensional parabolic bands with constant density of states, such that $N_{\alpha} = (|E^u_{\alpha}| - |E^l_{\alpha}|)^{-1}$, where $E^{l/u}_{\alpha}$ are the extrema of $\alpha$-band as shown in Fig.~\ref{toy_elhl}. The one-loop fermionic self-energy mediated via an interaction $V_{\alpha\beta}({\bf q},\Omega_m)$ is generally given  by
\begin{align}   \label{SE_appendix}  
     \Sigma_\alpha(\mathbf{k},\omega_m)&\notag\ \\
    = \frac{ T}{ N_q}&\displaystyle{\sum_{\mathbf{q},\Omega_{m}}} V_{\alpha \beta}(\mathbf{q},\Omega_{m}) G_{\beta}^0(\mathbf{k-q},\omega_{m}-\Omega_{m}),
\end{align}
where $ \Omega_m= 2m\pi T $ are the bosonic Matsubara frequencies and $\alpha,\beta$ are band indices. By approximating the full interaction with a  momentum-independent repulsive interband interaction $U( 1-\delta_{\alpha  \beta})D(\Omega_m)$, mediated via a bosonic propagator $D(\Omega_m)$ with a single Einstein mode at $\Omega_E$, and solving Eq.~(\ref{SE_appendix}) in the $T \rightarrow 0$ limit, one arrives at the analytical result discussed in Ref.~\onlinecite{Ortenzi2008}: 
\begin{align}  \label{SE_theory}
  \Sigma'_\alpha(\omega_0) = - \Omega_E U N_\beta \text{ln} \left| \frac{E_\beta^u }{E_\beta^l} \right|.
\end{align}
Notice that in the particle-hole symmetric limit where,  $|E_\beta^u|  = |E_\beta^l|$, Eq.~(\ref{SE_theory}) leads to $\Sigma'_\alpha(\omega_0)=0$. However, for particle-hole asymmetric bands,  $\Sigma'_\alpha(\omega_0)$ is non-zero. It has negative value for a hole-like band since $|E_{\tilde{\textnormal{M}}}^u| > |E_{\tilde{\textnormal{M}}}^l|$ and positive value for an electron band as $|E_\Gamma^l| > |E_\Gamma^u|$. This leads to simultaneous pocket shrinkage.

We now illustrate how the pocket shrinkage mechanism proposed by Ortenzi {\em et al.} works within a slightly more complex model with a momentum-dependent interaction.  Since the Fermi pockets in FeSC are small, it seems natural to associate zero-energy scattering processes at  small momentum transfer with intraband interactions, and those at momenta close to  the antiferromagnetic instability wavevector $(\pm \pi,0)$ with interband interactions. Here we discuss which region in momentum space is actually associated with interband processes that lead to pocket shrinkage.

As shown in Fig.~\ref{toy_elhl}, in the case of finite-energy interband scattering of a quasiparticle from the $\Gamma$ point with momentum transfer of ${\pi - \mathbf q_\mathrm{rad}}$, one can end up at an energy level of $E_{\tilde{\textnormal{M}}}({\tilde{\textnormal{M}}- \mathbf q_\mathrm{rad}})$. Similarly, scattering from $\tilde{\textnormal{M}}$ can end at an energy level of $E_\Gamma({\Gamma+ \mathbf q_\mathrm{rad}})$.  This means that for a  given interaction with momentum transfer ${\pi - \mathbf q_\mathrm{rad}}$, we need to perform the numerical integration over $\textbf{q}$ in the sum in Eq.~(\ref{SE_appendix}) using  a patch of radius ${\mathbf q_\mathrm{rad}}$ centered around $(\pm \pi,0)$. By changing the area of integration we are effectively changing the range of energy considered, which allows us to rewrite the analytical result of Eq.~(\ref{SE_theory}) as a function of ${\mathbf q_\mathrm{rad}}$,
\begin{align} \label{SE_analytic2}
  (\Sigma'_\Gamma)_\mathrm{an} & = - \Omega_E U N_{\tilde{\textnormal{M}}}  \text{ln} \left| \frac{E_{\tilde{\textnormal{M}}}({\tilde{\textnormal{M}}- \mathbf q_\mathrm{rad}}) }{E^l_{\tilde{\textnormal{M}}}} \right|, \\
  (\Sigma'_{\tilde{\textnormal{M}}})_\mathrm{an} & = - \Omega_E U N_\Gamma  \text{ln} \left| \frac{E^u_\Gamma}{E_\Gamma({\Gamma+\mathbf q_\mathrm{rad}}) } \right|,
  \label{SE_analytic1}
\end{align}
where the subscript ``an" stands for analytical. The sign of the self-energy renormalizations now crucially depends on the region of momenta used in the integration. A shrinking of the pockets is recovered as long as $|E_{\tilde{\textnormal{M}}}({\tilde{\textnormal{M}}- \mathbf q_\mathrm{rad}})| > |E_{\tilde{\textnormal{M}}}^l|$ and $|E_\Gamma({\Gamma+ \mathbf q_\mathrm{rad}})| > |E_\Gamma^u|$, while the opposite result is obtained if one restricts the integration over a small region of ${\bf q}$-space centered around $(\pm \pi,0)$-fluctuation.

We discuss this result in more detail in  Appendix \ref{RPA_num}, by performing numerical evaluations of Eq.~(\ref{SE_appendix}) as a function of ${\mathbf q_\mathrm{rad}}$ with the interaction $V_{\alpha \beta}(\mathbf{q},\Omega_{m})$ treated in the usual RPA approach. The main conclusion of our analysis of the toy-model is that the shrinking mechanism proposed by Ortenzi {\em et al.}  produces the correct trend for most band structures encountered in the FeSC. However, when attributing Fermi pocket shrinkage to interband scattering, though the interaction is sharply peaked around antiferromagnetic wavevector $(\pm \pi, 0)$, one needs to account for the crucial role played by all other momentum transfer $\mathbf{q}$-vectors participating in finite-energy scattering processes. Approximations that incorporate only large-$\bf q$ processes in an attempt to describe dominant interband scattering may fail to produce the correct pocket shrinkage, even if the interaction vertex is strongly peaked in $\bf q$.  Notice that the inclusion of the explicit momentum dependence of the spin spectrum in the above description of Fermi pocket shrinkage would lead to further renormalization, i.e., vertex corrections connected to the self-energy by standard Ward identities. We neglect such vertex corrections in the present work, though their effects could be important in the analysis of transport as shown, e.g., in Ref.~\onlinecite{fanfarillo_prl2012}. It is also worth noticing that using a simplified renormalization group analysis for FeSC, one can generate non-local interactions that can eventually lead to shrinkage or expansion of Fermi pockets~\cite{ChubukovFernandes2016}, the physical essence of which is similar to our findings.

\section{Model and RPA scheme} \label{model}

In the following sections, we perform numerical calculations for the realistic band structures of various FeSC. Based on the analysis of the simplified two-band model performed in  section~\ref{toy_model}, we expect to find a shrinkage of the FS in systems with shallow,  particle-hole asymmetric bands. However, the extent of this effect can be strongly modified by the number of hole/electron bands interacting via $V(\mathbf{q},\Omega_{m})$, the relative weight of the bands controlled by its density of states and the degree of particle-hole asymmetry of each band.

The realistic calculations of renormalized band structure in FeSC begin with a five-orbital tight binding Hamiltonian $H_0$. The kinetic energy $H_0$ includes the chemical potential $\mu_0$ and is spanned in the basis of five Fe $d$ orbitals $ [ d_{x y},d_{x^2 - y^2}, d_{x z},d_{y z},d_{3z^2 -r^2} ]$. The local interactions are included via Hubbard-Hund part $H_I$ and the full Hamiltonian is
    \begin{align} \label{Hamiltonian}
    \begin{split} 
     H   &= H_0 + H_I \\
        &= \sum_{ij\sigma} \sum_{qt} ( t_{ij}^{tq} - \mu_0\delta_{ij}\delta_{tq} ) c^\dagger_{it\sigma} c_{jq\sigma} \\
        & + U \sum_{it} n_{it\uparrow} n_{it\downarrow} + U' \sum_{i,t<q} \sum_{\sigma \sigma'} n_{it\sigma} n_{iq\sigma'} \\
        & + J \sum_{i,t<q} \sum_{\sigma \sigma'} c^\dagger_{it\sigma} c^\dagger_{iq\sigma'} c_{it\sigma'} c_{iq\sigma}  \\
        & + J' \sum_{i,t\neq q} c^\dagger_{it\uparrow} c^\dagger_{iq\downarrow} c_{it\downarrow} c_{iq\uparrow},
    \end{split}					
    \end{align}
where the interaction parameters $U,U',J,J'$ are given in
the notation of Kuroki \textit{et al.}~\cite{Kuroki2008}. We consider cases which obey spin-rotation invariance through the relations $U'=U-2J$ and $J=J'$. Here $q$ and $t$ are the orbital indices and $i$ is the Fe-atom site. The single-particle non-interacting Green's function is given by
    \begin{align} \label{Gbare}
     G^0(\mathbf{k},\omega_{m}) = \left[ i\omega_{m} - H_0(\mathbf{k}) \right]^{-1}.
    \end{align}
The orbitally resolved non-interacting susceptibility is
\begin{align} \label{baresus}
    & \chi^0_{pqst}(\mathbf{q},\Omega_m)\notag \\
    &= -\frac{1}{\beta N_k }  \sum_{\mathbf{k},\omega_m} G^0_{tq}(\mathbf{k},\omega_m) G^0_{ps}(\mathbf{k+q},\omega_m+\Omega_m) 	,
\end{align}
where $ N_k $ is the number of $\mathbf{k}$ points and $\beta=1/T$ is the inverse temperature.

Within the RPA, we define the charge-fluctuation and spin-fluctuation parts of the RPA susceptibility as
    \begin{align} \label{RPAsus}
    \begin{split} 
    \chi^C(\mathbf{q},\Omega_m) &=  [ 1 + \chi^0(\mathbf{q},\Omega_m) U^C]^{-1} \chi^0(\mathbf{q},\Omega_m) ,\\
    \chi^S(\mathbf{q},\Omega_m) &=  [ 1 - \chi^0(\mathbf{q},\Omega_m) U^S]^{-1} \chi^0(\mathbf{q},\Omega_m) .
    \end{split}					
    \end{align}
The interaction matrices $U^C$ and $U^S$ in orbital space have the following elements:
   \begin{align} \label{U_matrix}
   \begin{array}{ll}
	 U^{C}_{pppp}= U     &  ,U^{S}_{pppp}= U  \\
	 U^{C}_{ppss}= 2U'-J &  ,U^{S}_{ppss}= J  \\
	 U^{C}_{pssp}= J'    &  ,U^{S}_{pssp}= J' \\
	 U^{C}_{psps}= 2J-U' &  ,U^{S}_{psps}= U'. \\
   \end{array}
   \end{align}
We note that our calculations are based on the local density approximation (LDA) \textit{ab-initio} electronic band structure which already contains important Hartree-Fock corrections. The remaining leading order spin-fluctuation contributions to the self-energy within RPA is mediated through the particle-hole interaction
 \begin{align} \label{Veqn}
 \begin{split}
     V_{pqst}(\mathbf{q},\Omega_{m}) &= \left[ \frac{3}{2} U^S \chi^S(\mathbf{q},\Omega_{m}) U^S + \frac{1}{2} U^C \chi^C(\mathbf{q},\Omega_{m}) U^C \right. \\
     &- \left. \left(\frac{U^C+U^S}{2}\right) \chi^0(\mathbf{q},\Omega_{m}) \left(\frac{U^C+U^S}{2}\right) \right]_{pqst}.
 \end{split}
 \end{align}
In the paramagnetic state, where the initial and the final spins for the normal self-energy are the same, the above interaction results from summing over all the possible triplet and singlet scattering channels~\cite{KemperPRB2011}. Using Dyson's equation, one obtains the interacting Green's function
\begin{align} \label{Gfull}
    G(\mathbf{k},\omega_{m})^{-1} = G^0 (\mathbf{k},\omega_{m})^{-1} - \Sigma(\mathbf{k},\omega_m),
\end{align}
where the self-energy in orbital basis is given by
    \begin{align}   \label{selfenergy}  
     \Sigma_{ps}(\mathbf{k},\omega_m)&\notag\ \\
    = \frac{1}{\beta N_q}&\displaystyle{\sum_{\mathbf{q},\Omega_{m}}\sum_{qt}} V_{pqst}(\mathbf{q},\Omega_{m}) G_{qt}^0(\mathbf{k-q},\omega_{m}-\Omega_{m}).
    \end{align}
To ensure particle number conservation, the chemical potential is reevaluated from the interacting Green's function (details in Appendix \ref{numeric_details}).

The scattering lifetime of quasiparticles at a specific momentum point $\mathbf{k}$ with energy $\omega$ in orbital $p$ is given by $-Z_p(\mathbf{k}) \Sigma^"_{pp}(\mathbf{k},\omega)$ , where $\Sigma^"$ refers to the imaginary part of the self-energy, and the quasiparticle weight $Z_p(\mathbf{k})$ is given by
\begin{align} \label{Zeqn}
    Z_p(\mathbf{k}) = \left[ 1-  \left. \frac{ \partial\Sigma^"_{pp}(\mathbf{k},\omega_m)}{\partial\omega_m} \right|_{i\omega_m \rightarrow 0^+ } \right]^{-1}
\end{align}
At sufficiently low temperatures, one can use the analytic properties of the Matsubara self-energy to approximate $Z_p(\mathbf{k}) \approx [ 1- \Sigma^"_{pp}(\mathbf{k},\omega_0)/\omega_0 ]^{-1}$, where $\omega_0=\pi T$. We set $T = 100\textnormal{K} = 0.0086$ eV for the rest of this paper unless mentioned otherwise. We have used standard $U$ and $J$ values as used in the literature employing the RPA approach~\cite{Graser2009}. We use the Pad\'e approximation for numerical analytic continuation. Further numerical details are provided in Appendix \ref{numeric_details}. 

\begin{figure*}[hbt!] 
     \centering
     \begin{tabular}{l}
          \subfloat[]{\includegraphics[width=0.475\textwidth]{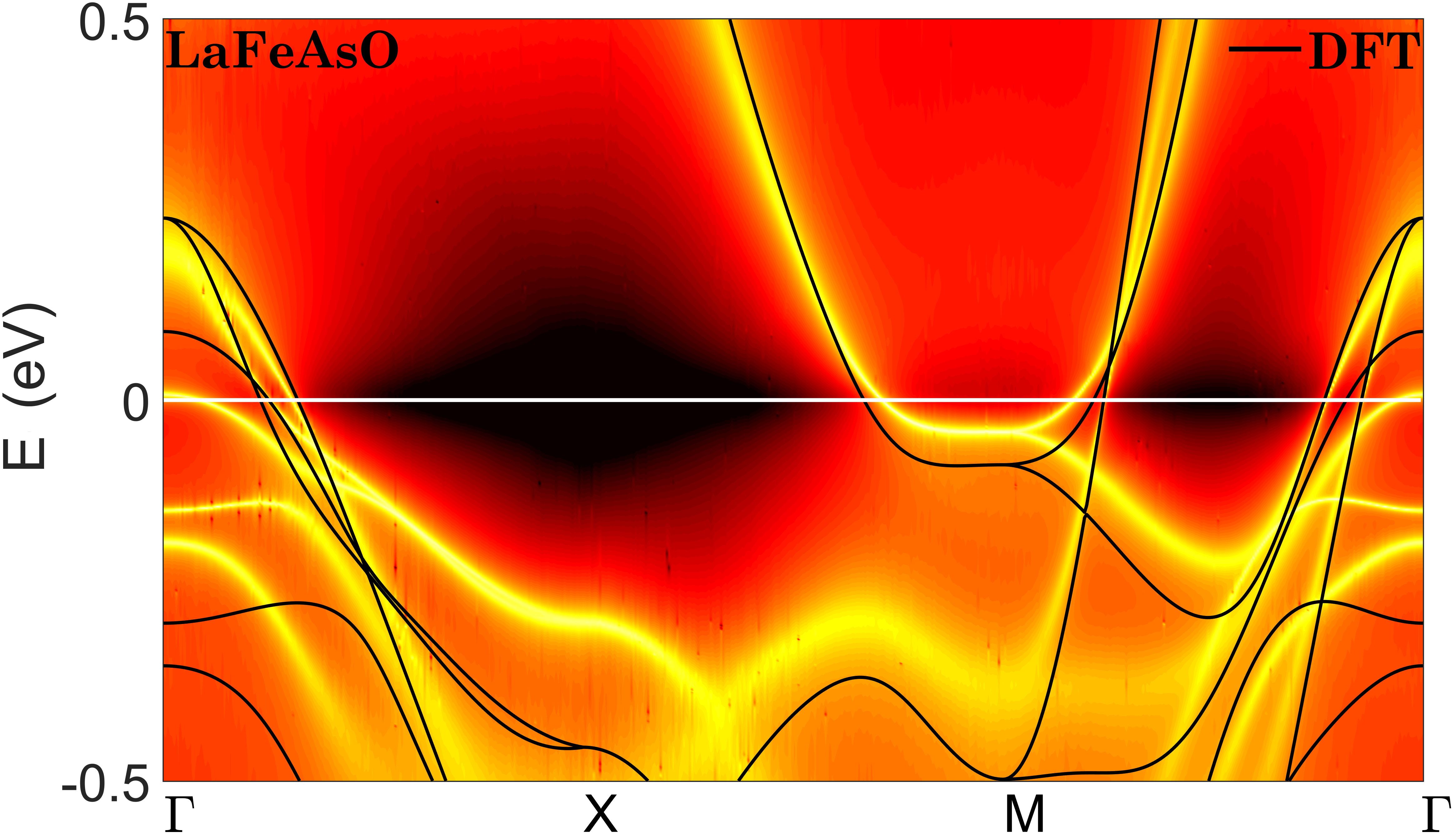}\label{1111_bands}} \\
          \subfloat[]{\includegraphics[width=0.475\textwidth]{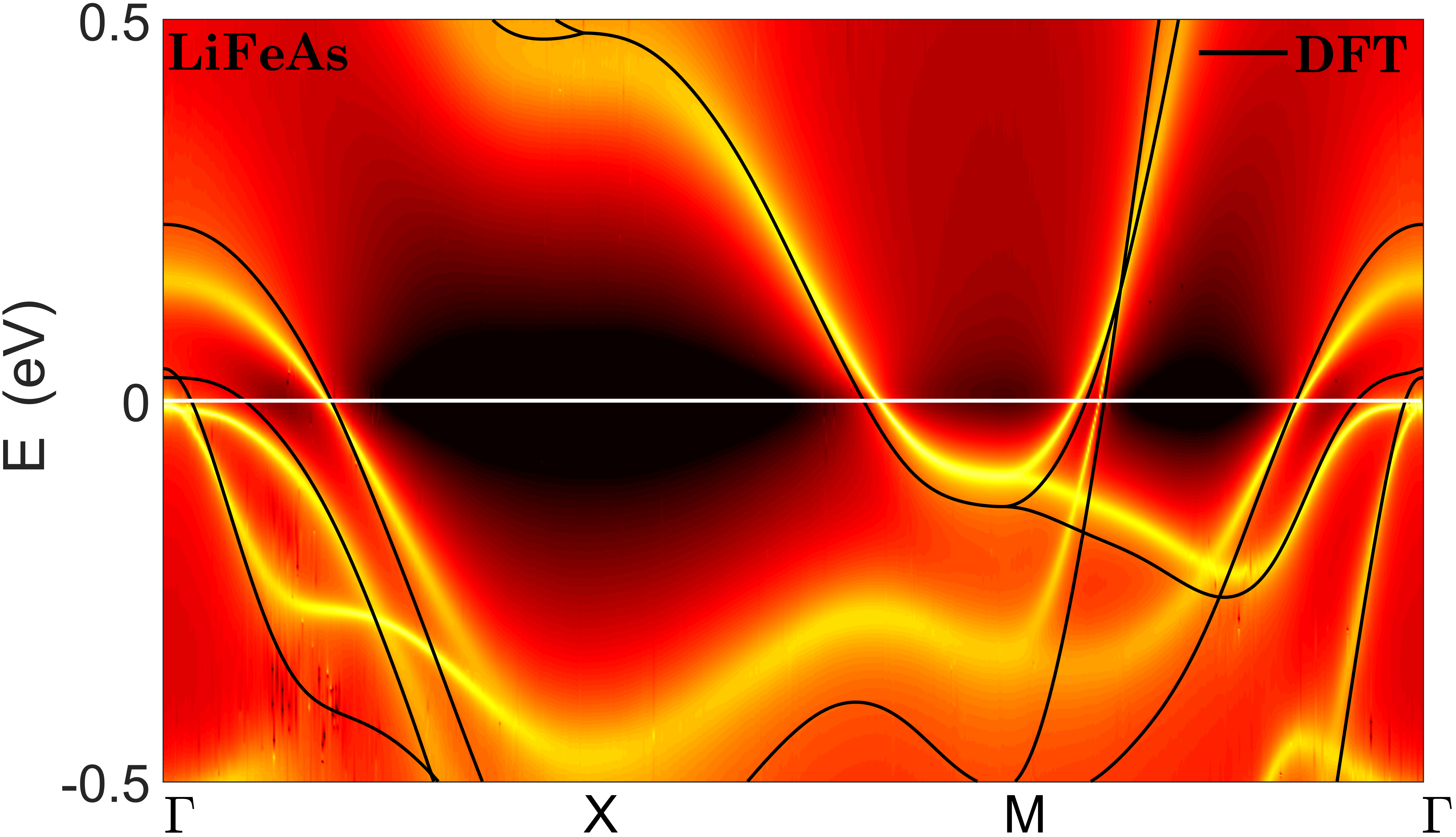}\label{111_bands}} \\
          \subfloat[]{\includegraphics[width=0.475\textwidth]{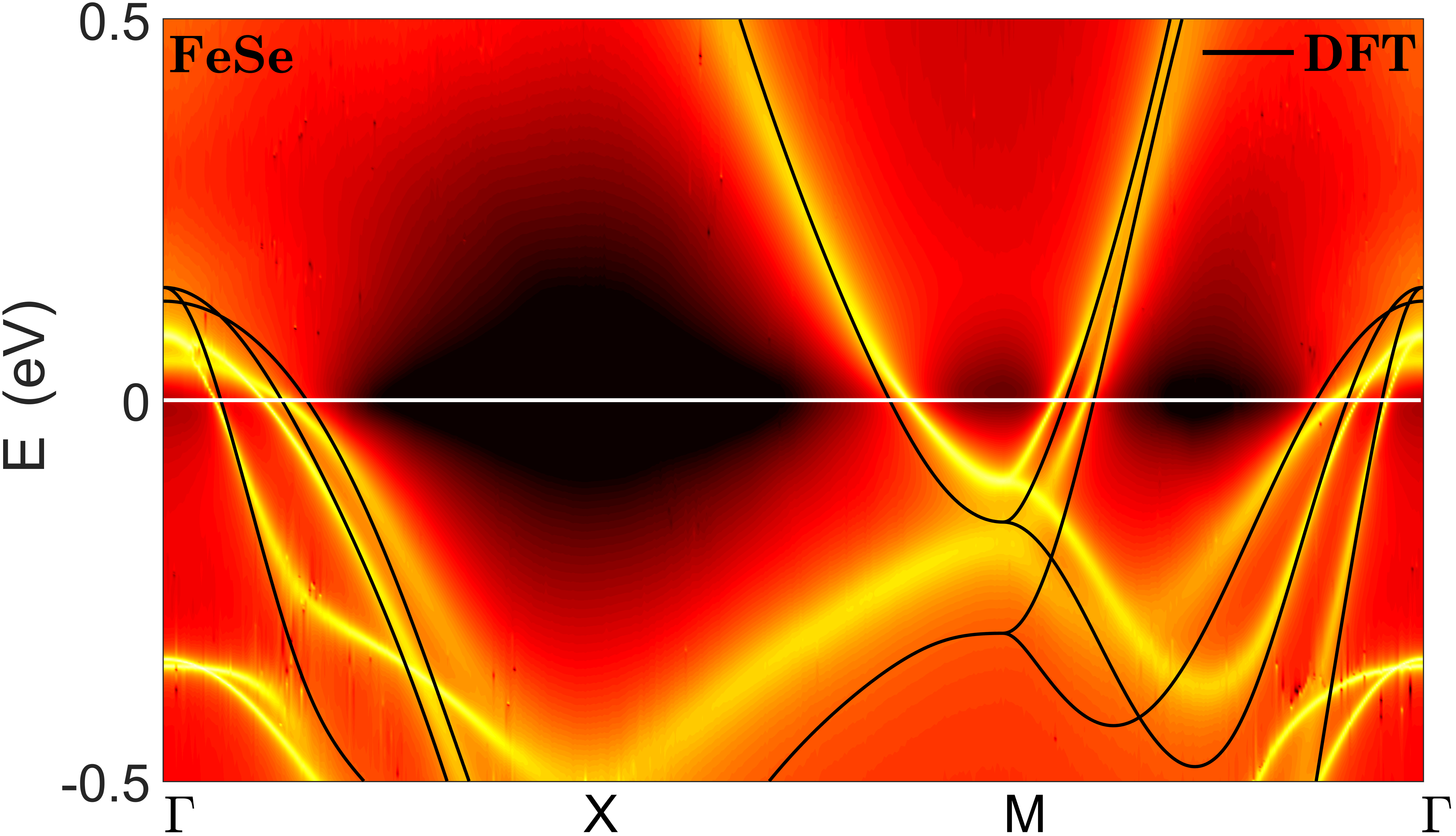}\label{11_bands}} 
     \end{tabular}
     \begin{tabular}{l}
          \subfloat[]{\includegraphics[width=0.265\textwidth]{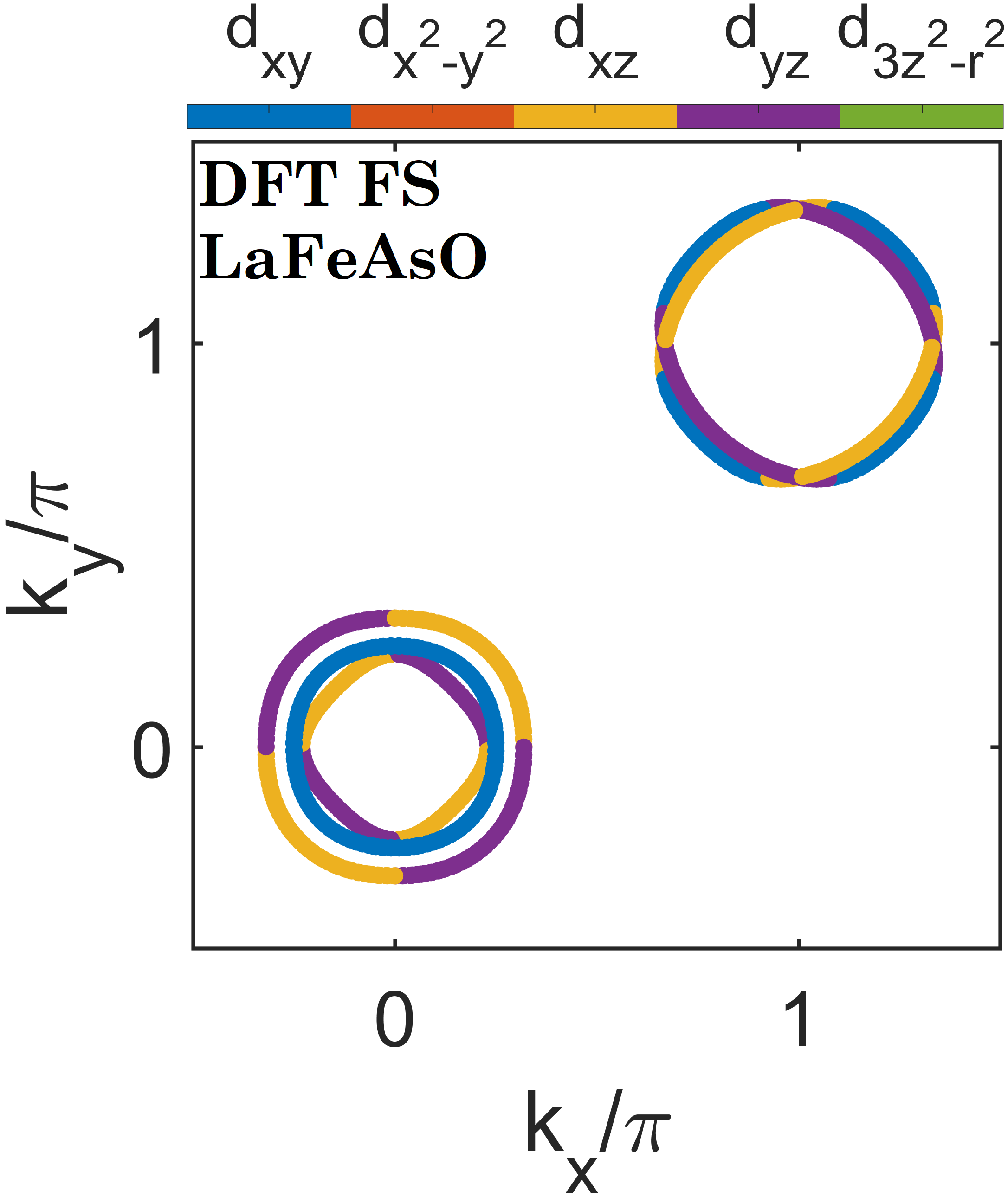}\label{1111_FS}} \\
          \subfloat[]{\includegraphics[width=0.265\textwidth]{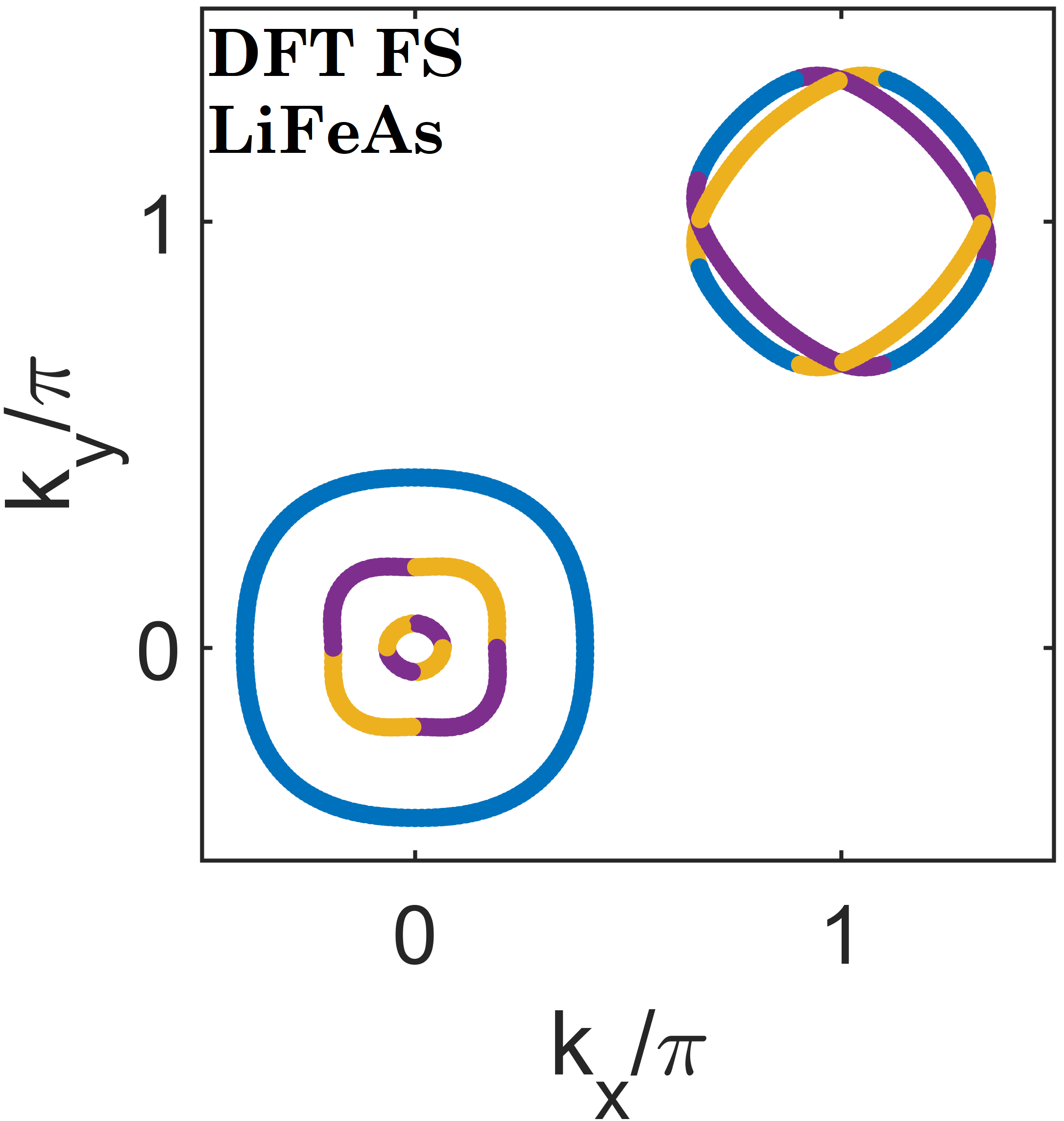}\label{111_FS}} \\
          \subfloat[]{\includegraphics[width=0.266\textwidth]{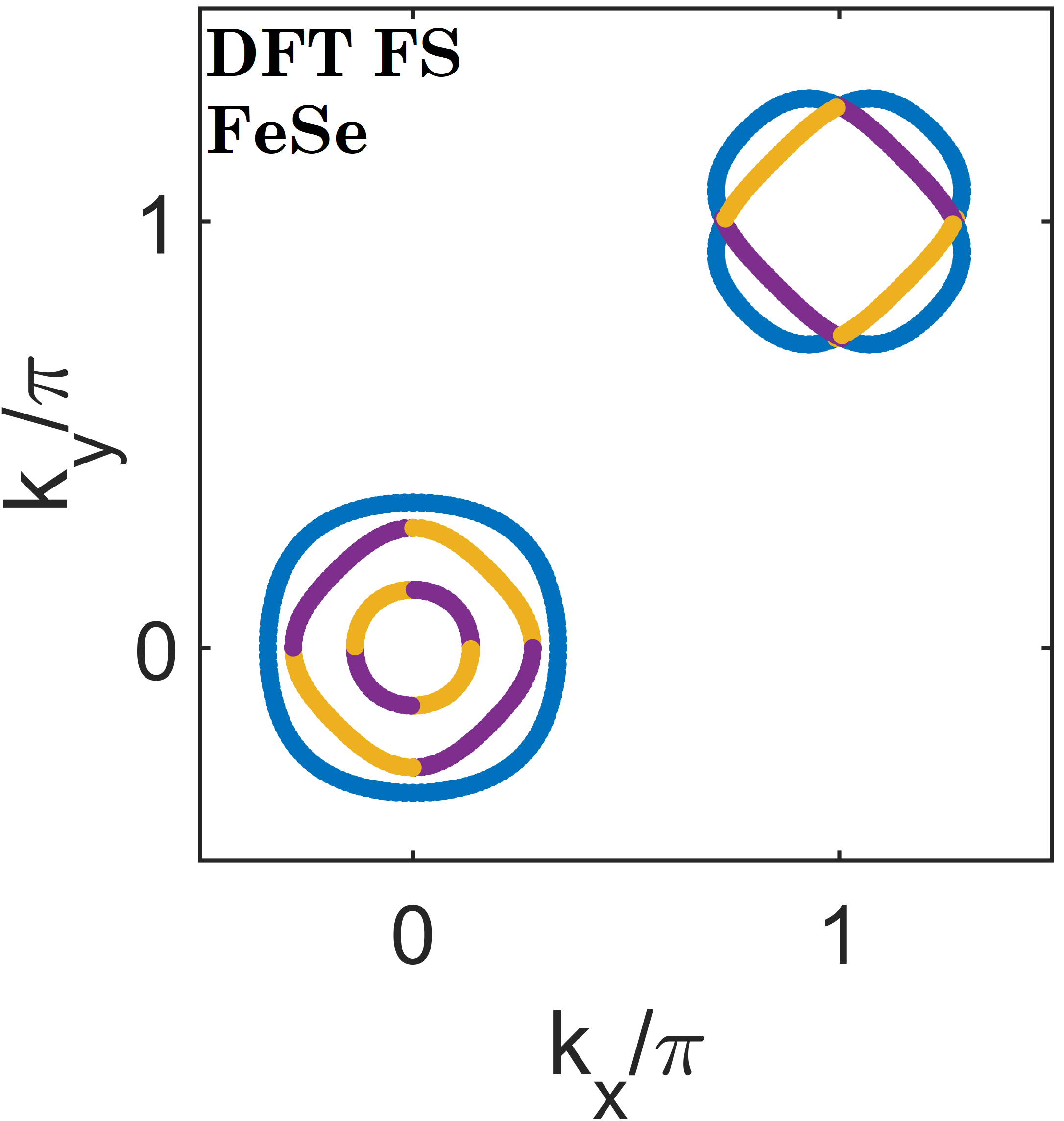}\label{11_FS}}
     \end{tabular}
     \begin{tabular}{l}
          \subfloat[]{\includegraphics[width=0.219\textwidth]{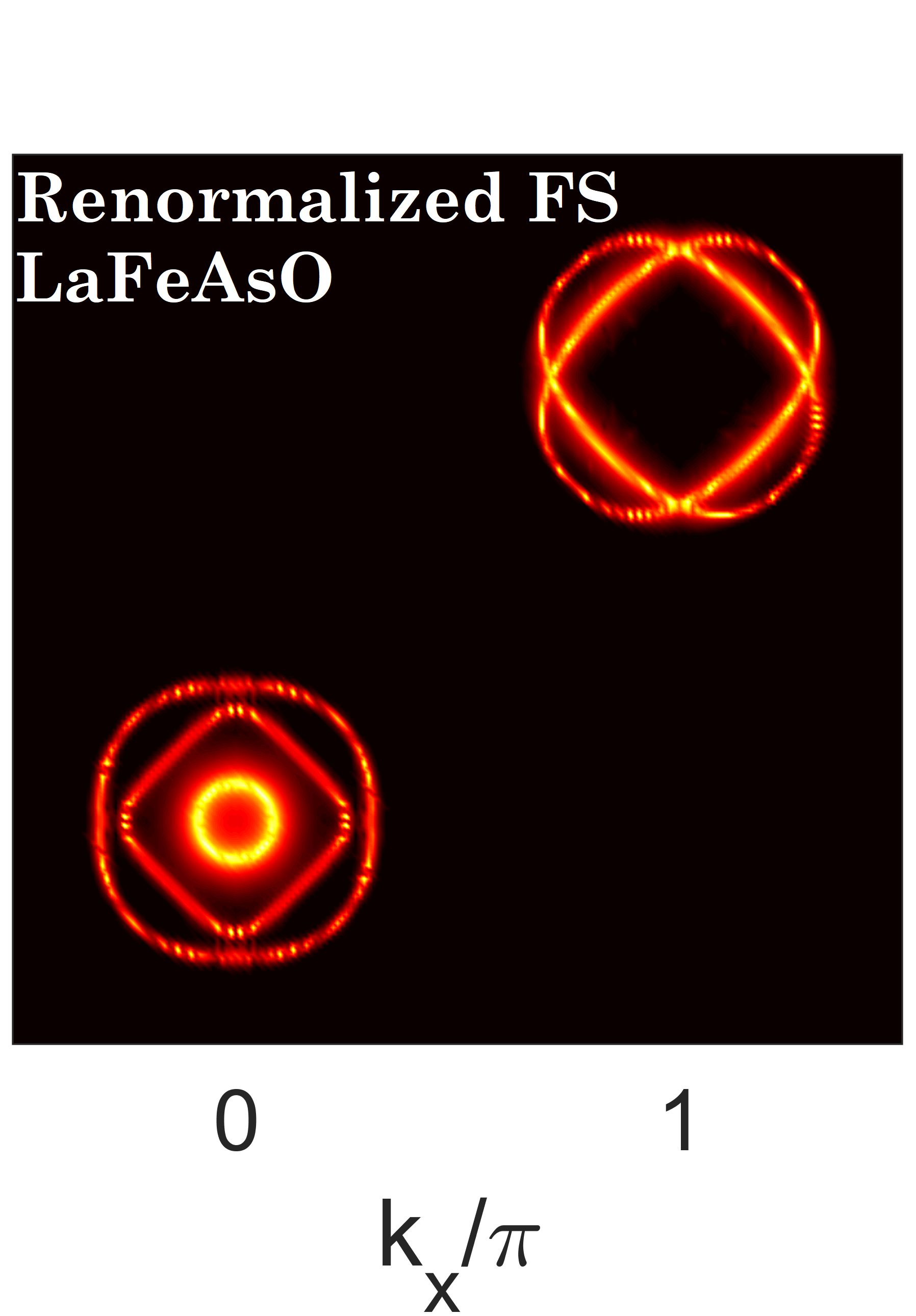}\label{1111_2Fe}} \\
          \subfloat[]{\includegraphics[width=0.217\textwidth]{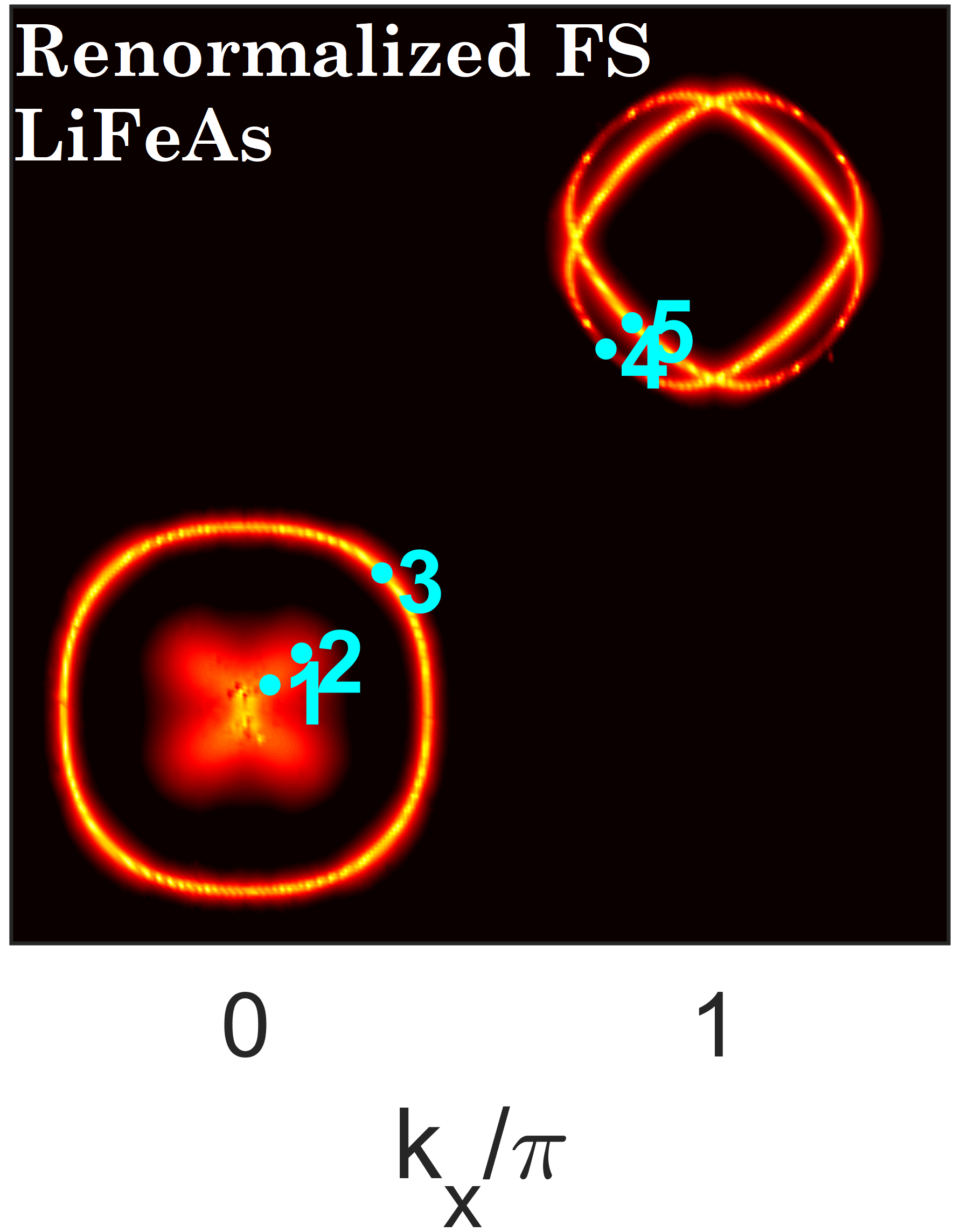}\label{111_2Fe}} \\
          \subfloat[]{\includegraphics[width=0.217\textwidth]{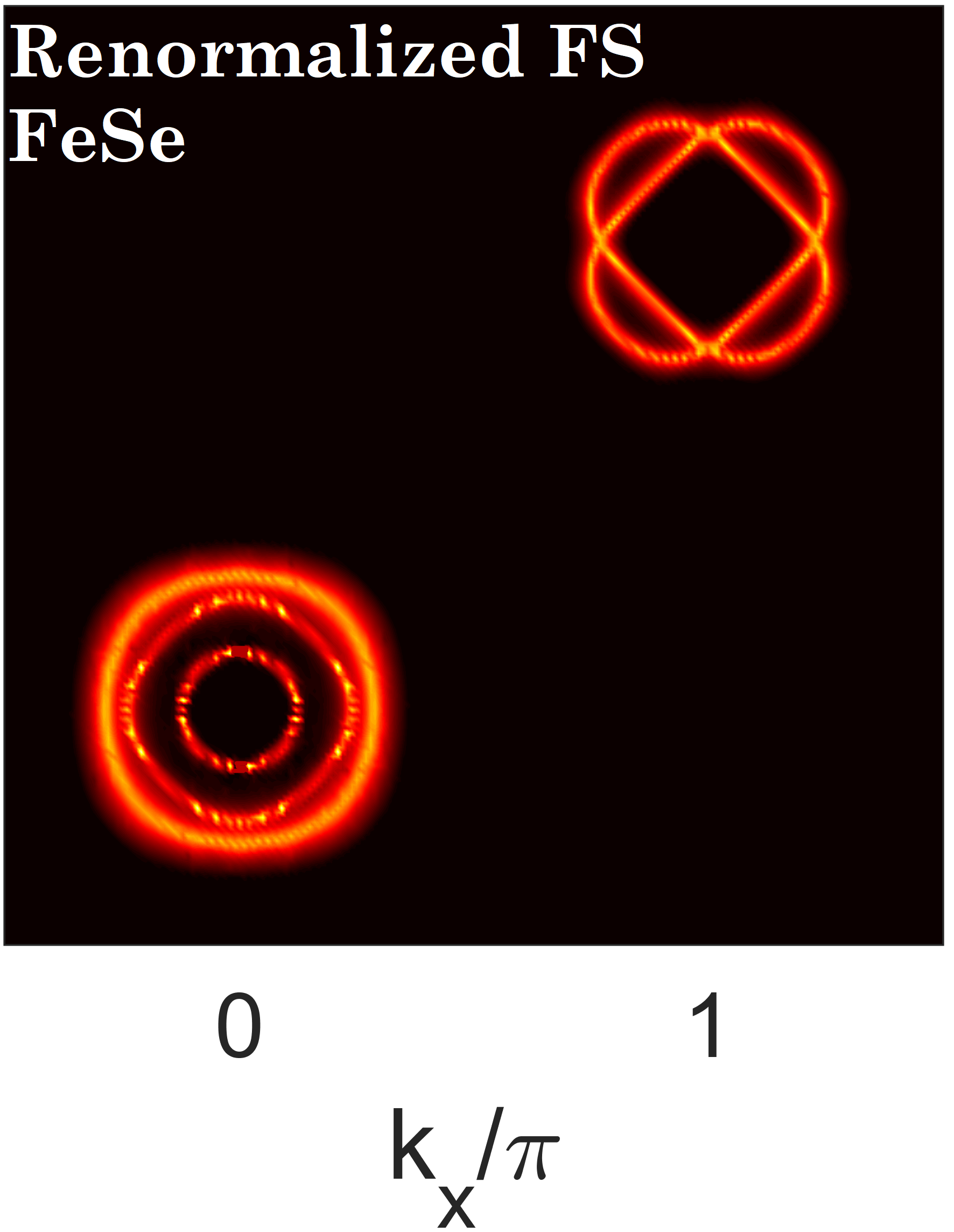}\label{11_2Fe}}
     \end{tabular}
	\caption{Spectral function $A(\mathbf{k},\omega)$ for interacting bands evaluated along high-symmetry path $\Gamma-X-M-\Gamma$ for (a) LaFeAsO at $U=0.8$ eV, (b) LiFeAs at $U=0.8$ eV, and (c) FeSe at $U=0.89$ eV in the $2$-Fe BZ. The value of Hund's coupling term was fixed at $J=U/8$. The black lines and the bright yellow curves denote the DFT and the renormalized bands, respectively. Contour plots of the DFT Fermi surface for (d) LaFeAsO, (e) LiFeAs, and (f) FeSe are represented by its corresponding dominant orbital character as indicated in the color legend on top. Spectral map of the renormalized Fermi surface for (g) LaFeAsO, (h) LiFeAs, and (i) FeSe. The momentum points marked as $1-5$ on LiFeAs Fermi surface in (h) are used in evaluation of scattering lifetimes of quasiparticles later.}
     \label{main_figs}
\end{figure*}

\section{Results}

\subsection{Electronic structure renormalization} \label{bs_results}

To present our results in a manner that is easily comparable to experimental data, we folded our spectral function from 1-Fe to 2-Fe BZ. As shown earlier in 
Fig.~\ref{schematic_cartoons}, the high-symmetry points in the 2-Fe BZ are $\Gamma = (0,0)$, $X = (\pi,0)$ and $M= (\pi,\pi)$. In Fig.~\ref{main_figs}, we show the spectral function $A(\mathbf{k},\omega) = -\frac{1}{\pi} \sum_p G^"_{pp}(\mathbf{k},\omega)$, obtained from the imaginary part of the retarded interacting Green's function within the RPA scheme for LaFeAsO, LiFeAs and FeSe in their tetragonal phase as representatives of the pnictide and chalcogenide classes. For our DFT electronic dispersion, we have chosen the  tight-binding parameters derived from DFT, relevant to LaFeAsO as in Ref.~\onlinecite{Graser2009}, LiFeAs as in Ref.~\onlinecite{ZantoutPRL2019}, and FeSe as in Ref.~\onlinecite{Eschrig09}. Due to the quasi-2D nature of FeSC bands and to compare with available experimental data, our evaluations are restricted to the 2D plane at $k_z =0$. The obtained quasiparticle bands closely trace the position of the peak in the spectral density, as denoted by the bright yellow curves in the spectral function [Fig.~\subref*{1111_bands} - LaFeAsO, \subref*{111_bands} - LiFeAs, \subref*{11_bands} - FeSe]. For easier comparison to the DFT bands, we have superimposed the DFT-derived band structure along high-symmetry path $\Gamma-X-M-\Gamma$ with black lines in these figures. The corresponding DFT FSs with their respective orbital content are shown in Fig.~\subref*{1111_FS} for LaFeAsO, \subref*{111_FS} for LiFeAs, and \subref*{11_FS} for FeSe. The RPA spectral maps for the renormalized FSs are provided in Fig.~\subref*{1111_2Fe} for LaFeAsO, \subref*{111_2Fe} for LiFeAs, and \subref*{11_2Fe} for FeSe. All results shown in Fig.~\ref{main_figs} evaluated within RPA are in agreement with findings from TPSC and FLEX. 

First of all, we notice the following universal trends across all the families of FeSC: the quasiparticle bands are strongly renormalized, resulting in noticeable changes in the FS, i.e., shrinkage around $\Gamma$ and $M$ points for both hole and electron pockets compared to their DFT counterparts, and momentum-dependent renormalization of the velocity of the bands under interaction. Upon comparison to experiment, we find that our renormalized FS for LaFeAsO [Fig.~\subref*{1111_2Fe}] are similar to the ones observed in ARPES~\cite{LaFeAsO_Arpes_PRB2010}. Out of the three $\Gamma$-centered hole pockets, the $d_{xy}$ pocket undergoes the most shrinkage while for the
$M$-centered electron pockets, the $d_{xz/yz}$ parts shrink more than the $d_{xy}$ part. For LiFeAs, the renormalized FS  in Fig.~\subref*{111_2Fe} agrees
with ARPES findings from Ref.~\onlinecite{BorisenkoLiFeAs,Lee_etal_Kotliar2012,Brouet_PRB_2016,Fink2019}. The $\Gamma$-centered $d_{xy}$ hole pocket does not show noticeable shrinkage but the inner-most $d_{xz/yz}$ hole pockets gets pushed down and show only residual spectral weight from the bands grazing the Fermi level. The renormalized FS for FeSe [Fig.~\subref*{11_2Fe}] also shows simultaneous shrinkage
of both hole and electron pockets, however, it disagrees  with  ARPES data~\cite{Coldea2016,Watson2016,Coldea_2017_review,Reiss2017,Watson_hubbard2017}, where drastic shrinkage of all the Fermi pockets is reported, including
vanishing of the $d_{xy}$ hole pocket from the $\Gamma$ point. The most striking aspect of the band renormalization, i.e, the complete disappearance of
the largest ($d_{xy}$) hole pocket found in DFT calculations based on reported X-ray crystal structures, has not been addressed in the literature.

As explained in Section \ref{toy_model} above, a momentum- and orbital-dependent self-energy that has a negative (positive) value of $\Sigma'(\mathbf{k},\omega=0)$ around the hole (electron) pocket is a consequence of itinerant spin-fluctuations and can result in pocket shrinkage to various degrees in different orbitals. It should be noted that the above calculations are based on DFT-derived tight-binding models without spin orbit coupling. Therefore, certain low-energy splitting have been neglected.  In particular, in the case of LiFeAs, it is known that only a single band crosses the Fermi level at the Z point, and neither of the hole bands at the $\Gamma$ point\cite{Wang13,Borisenko2018}. Inclusion of spin-orbit coupling can account for this splitting and will allow for agreement with ARPES data only at somewhat higher values of the Coulomb interaction.

\subsection{Effects of nearest-neighbor Coulomb repulsion} \label{longrangeCoulomb}

\begin{figure}[bt!]
    \centering
    \includegraphics[width =\linewidth]{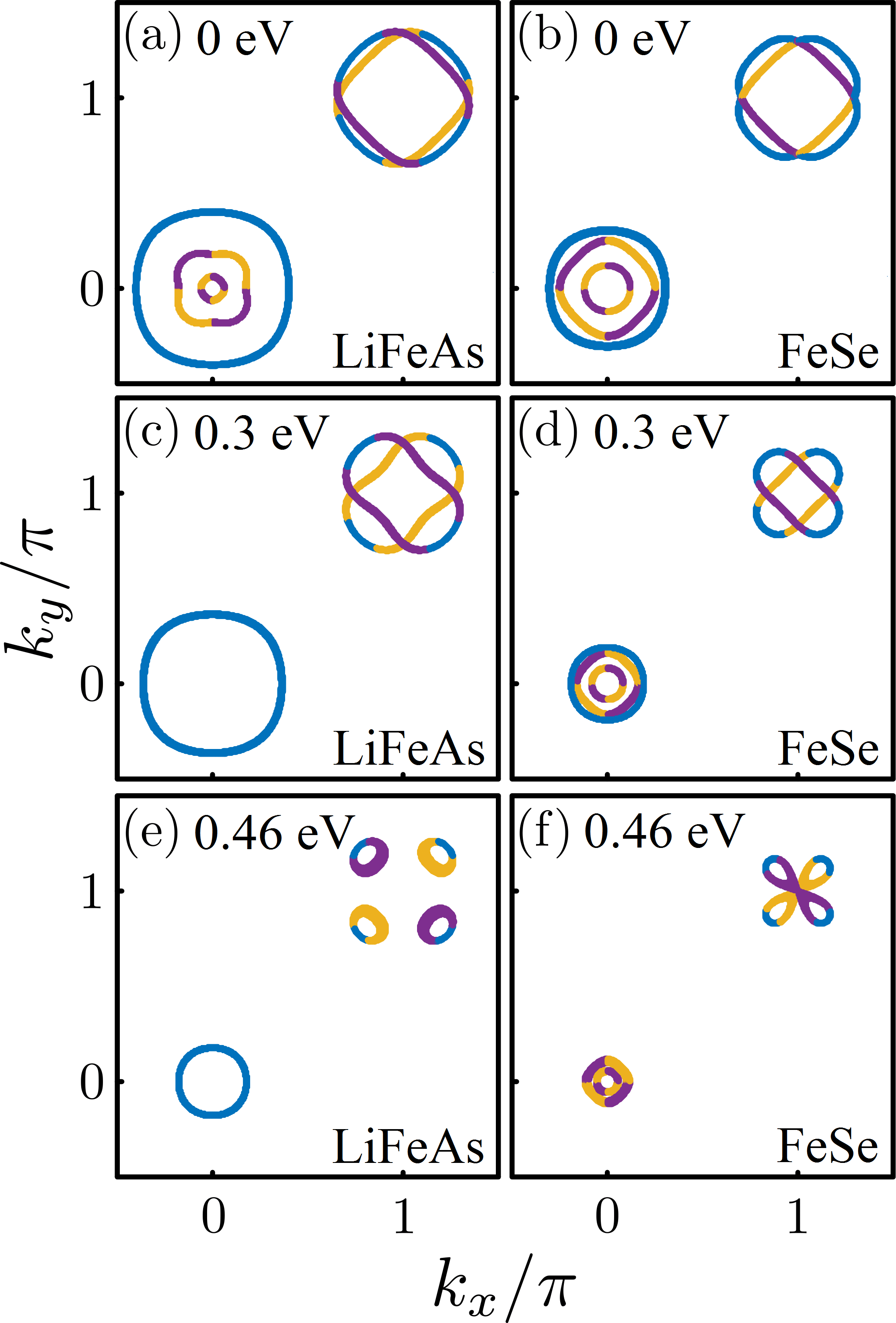}
    \caption{The effects of NN Coulomb interactions on the Fermi surfaces of LiFeAs and FeSe. (a,c,e) Fermi surface for LiFeAs with the interaction potential strength $V=0,\,0.30,\,0.46$ eV, respectively. (b,d,f) Same as in (a,c,e), but for FeSe. All figures were obtained using $T=0.01$ eV and a filling of $n=6$.}
    \label{fig:FSs_FeSe_LiFeAs_br}
\end{figure}

\begin{figure*}[hbt!]
    \centering
    \subfloat[]{\includegraphics[width=0.31\textwidth]{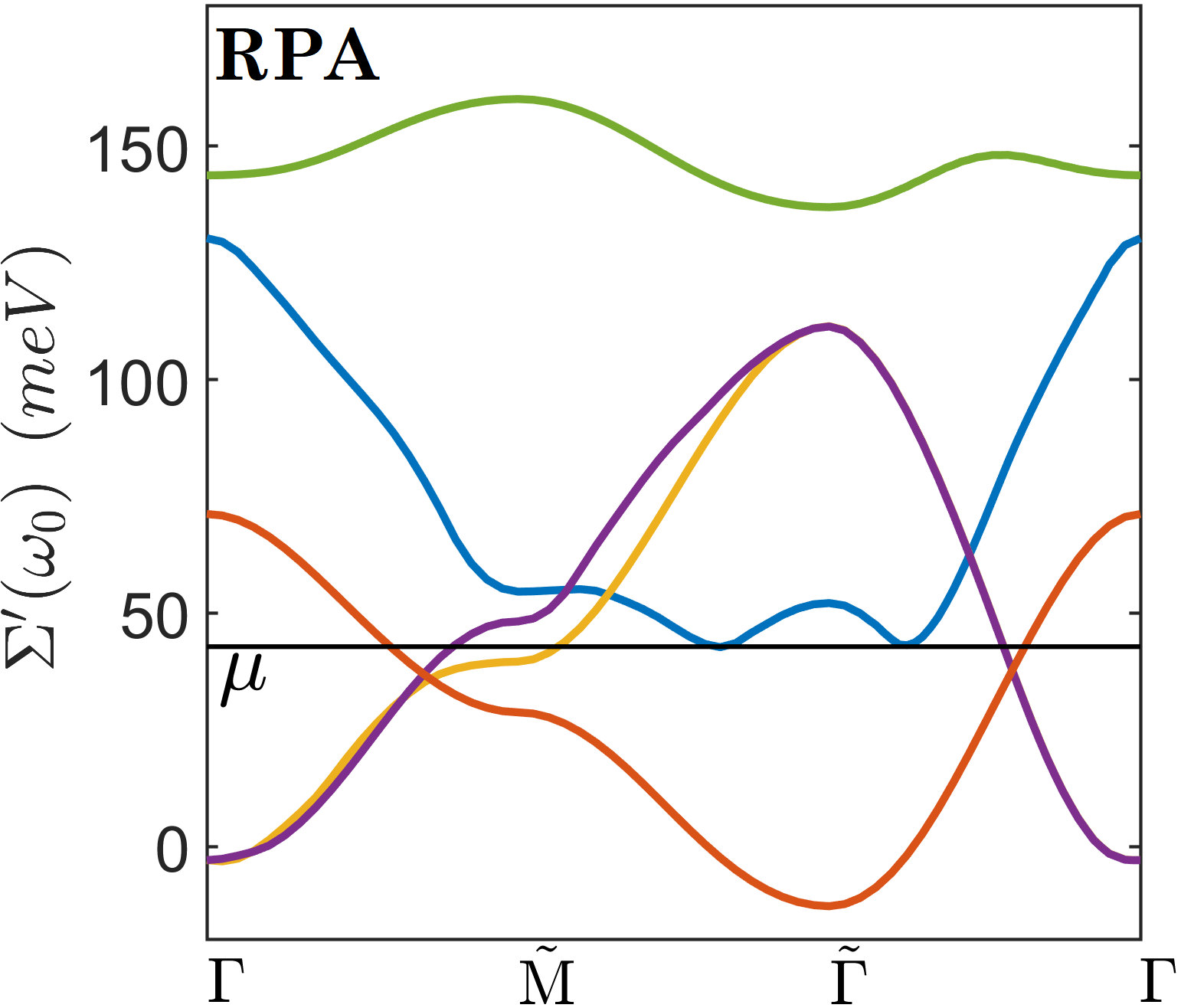}\label{RPA_111}} \;
    \subfloat[]{\includegraphics[width=0.316\textwidth]{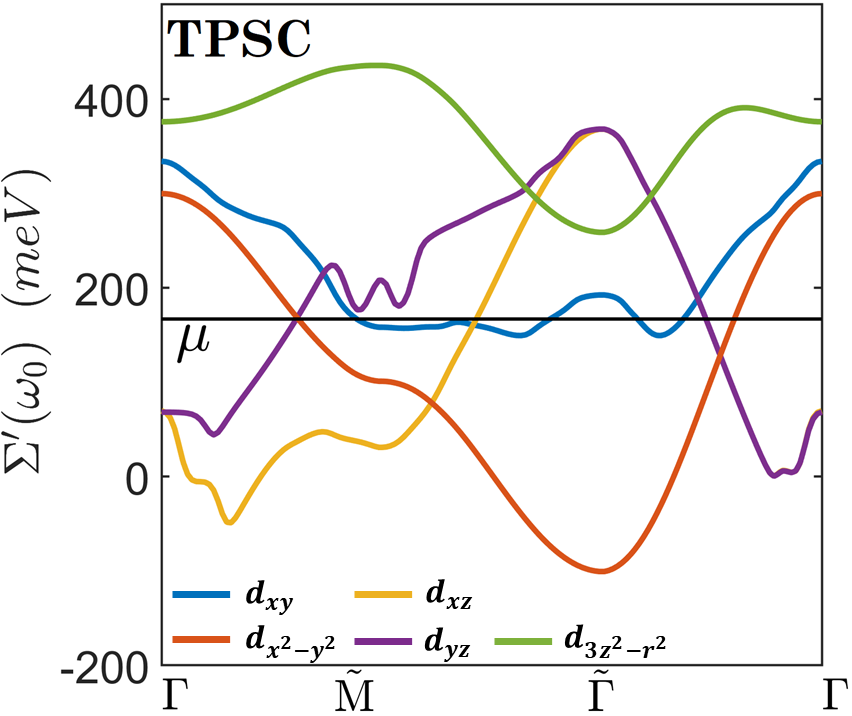}\label{TPSC_111}} \;
    \subfloat[]{\includegraphics[width=0.314\textwidth]{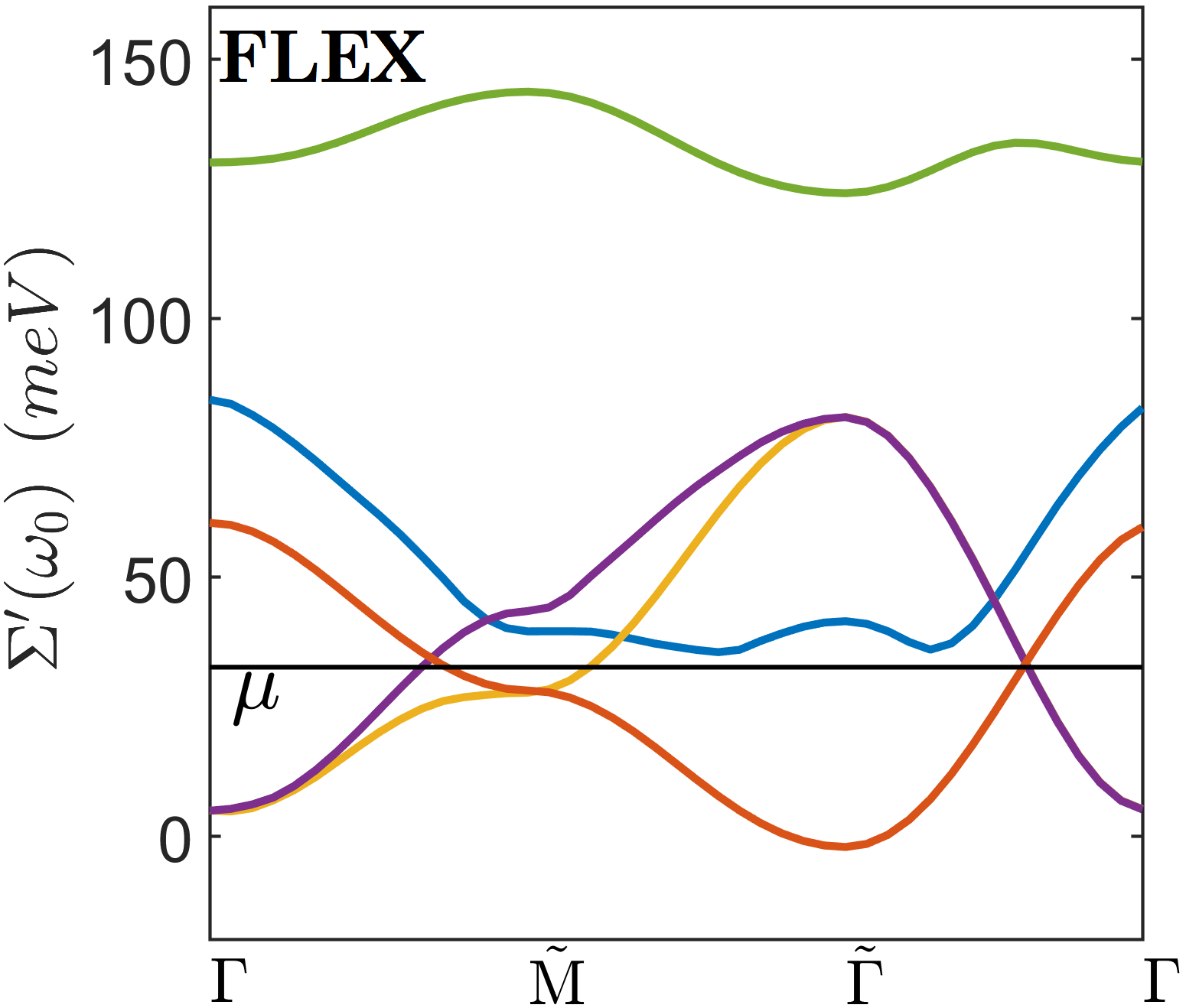}\label{FLEX_111}}
    \caption{Real part of the orbitally-resolved static self-energy $\Sigma'(\mathbf{k},\omega_0)$ computed along high-symmetry path $\Gamma-\tilde{\textnormal{M}}-\tilde{\Gamma}-\Gamma$ in the 1-Fe BZ of LiFeAs via (a) RPA, (b) TPSC, and (c) FLEX evaluation scheme. The orbital character and its corresponding color choice is same as before as marked in the color legend at the bottom of Fig (b). The solid black line demarcates the value of the new chemical potential $\mu$. The RPA and FLEX evaluations were carried out at $U=0.8$ eV and $J=U/8$. The TPSC parameters used are the same as in Ref. \onlinecite{ZantoutPRL2019}.}
    \label{compare_plots}
\end{figure*}

In this section we explore another physical mechanism that may be relevant for a quantitative understanding of band renormalizations. Longer-range Coulomb interactions have been previously shown to influence the pairing structure in FeSC~\cite{ZWang2011}, and proposed as a candidate for the origin of $d$-wave nematic order in FeSe~\cite{Yi_2015,Jiang2016Mar}.
These interactions can also result in a drastically renormalized band structure~\cite{Yi_2015,Jiang2016Mar,Scherer2017Mar}. In order to explore this effect, for the  electronic structures of LiFeAs and FeSe considered here, we introduce the following nearest-neighbor (NN) Coulomb interaction
\begin{equation}
    H_{V} = V\sum_{\langle i, j \rangle} \sum_{t < q} \sum_{\sigma,\sigma'} n_{i t \sigma} n_{j q \sigma'},
\end{equation}
where $V > 0$, and $\langle i, j \rangle$ indicates the summation over NN sites. To treat the effects of the NN Coulomb repulsion, we perform a Hartree-Fock mean-field decoupling and introduce the bond-order fields
\begin{align}
    &N_{\mathbf{k} \sigma}^{t q} = 
    \\
    &-\frac{2V}{N_{k}} \sum_{\mathbf{k}'} [\cos(k_x - k_x') + \cos(k_y - k_y')]\langle c^{\dagger}_{\mathbf{k}' q \sigma} c^{ }_{\mathbf{k}' t \sigma} \rangle.\nonumber
\end{align}
Note that here we investigate only the effects of NN coupling on the bandstructure. It remains to be determined what are the effects of further next-nearest-neighbor interactions. In addition, we ignore Hartee-Fock terms arising from onsite interactions since these only introduce orbital-dependent onsite energy shifts that, we find for interaction parameters corresponding to Fig. \ref{main_figs}, lead to small quantitative changes to the results discussed below.  A quantitative exploration of the effects of a full set of neighboring Coulomb interaction parameters is left to a future study.

By closely following the symmetry arguments in Refs. \onlinecite{Yi_2015,Jiang2016Mar,Scherer2017Mar}, we decompose the bond-order fields into C$_4$-symmetry-preserving and -breaking terms, denoted $N_{\mathbf{k}\sigma, {\rm br}}^{tq}$ and $N_{\mathbf{k}\sigma, {\rm sb}}^{tq}$, respectively. The additional subscripts of the fields refer to band renormalization (br) and symmetry-breaking (sb). For our current study, we assume no spontaneous structural transition to occur, and therefore neglect the C$_4$-symmetry-breaking terms. This leaves us with the task of calculating $N_{\mathbf{k} \sigma, {\rm br}}^{tq}$ self-consistently, while keeping the electron density $n$ fixed. The resulting FSs for LiFeAs and FeSe are depicted in Fig.~\ref{fig:FSs_FeSe_LiFeAs_br} for different interaction potential strengths $V$ as used in the standard literature~\cite{Scherer2017Mar}. 
It is evident that the FSs are significantly modified even for rather small $V$, and that the two systems are affected quite differently by the NN Coulomb repulsion. These observations are seen specifically for $V=0.30$ eV
[Fig.~\ref{fig:FSs_FeSe_LiFeAs_br}{\blue (c)} and \ref{fig:FSs_FeSe_LiFeAs_br}{\blue (e)}], where the $d_{xz/yz}$-dominated $\Gamma$-pockets of the LiFeAs band get completely removed from the Fermi level, whereas for the FeSe band in Fig.~\ref{fig:FSs_FeSe_LiFeAs_br}{\blue (d)} and \ref{fig:FSs_FeSe_LiFeAs_br}{\blue (f)}, it is the $d_{xy}$-dominated $\Gamma$-pocket that gets removed for $V=0.46$ eV while the $d_{xz/yz}$-dominated pockets prevail. The origin of this difference between the two materials can be traced to the initial band structures [see Fig.~\subref*{111_bands} and \subref*{11_bands}]. Specifically, the $d_{xy}$-dominated DFT hole band at $\Gamma$ reaches significantly larger energies for LiFeAs ($\approx 250$ meV) than for FeSe ($\approx 100$ meV), thus simply making it more robust towards renormalization effects.

\subsection{Comparison of RPA, TPSC and FLEX results} \label{comparison} 

In order to provide a quantitative basis of comparison of the self-energy obtained via different methodologies described in this paper, we  focus on the real part of the orbitally-resolved static self-energy $\Sigma'(\mathbf{k},\omega_0)$ computed along high-symmetry path $\Gamma-\tilde{\textnormal{M}}-\tilde{\Gamma}-\Gamma$ in the 1-Fe BZ of LiFeAs. $\Gamma$ hosts the small $d_{xz/yz}$-dominated hole pockets. $\tilde{\Gamma}$ refers to the $(\pi,\pi)$ point in the 1-Fe BZ which gets folded back to the $\Gamma$ point in the 2-Fe BZ representation, and hosts the large $d_{xy}$-dominated hole pocket. $\tilde{\textnormal{M}}$ hosts the $d_{yz/xy}$-dominated electron pocket which gets folded and represented by the $M$-point in the 2-Fe BZ. In Fig.~\ref{compare_plots}, we plot $\Sigma'(\mathbf{k},\omega_0)$ along $\Gamma-\tilde{\textnormal{M}}-\tilde{\Gamma}-\Gamma$ obtained via RPA (\subref*{RPA_111}), TPSC (\subref*{TPSC_111}) and FLEX evaluations (\subref*{FLEX_111}). 

We have demarcated the new chemical potential $\mu$ by the solid black line to illustrate the relative sign contrast of $\Sigma'(\mathbf{k},\omega_0)$ between different momentum points. First, we notice the qualitative similarities in $\Sigma'(\mathbf{k},\omega_0)$ obtained via the three methods. The $d_{xz/yz}$ component at the $\Gamma$ point lies below the new chemical potential $\mu$ resulting in a relative negative value of the self-energy, whereas the $d_{yz/xy}$ component at the $\tilde{\textnormal{M}}$ point lying above $\mu$, results in a relative positive value. As discussed before in Section \ref{toy_model}, due to the above behavior of the self-energy, the resulting renormalized Fermi surface shows pocket shrinkage for both the inner $d_{xz/yz}$ hole and $d_{yz/xy}$ electron pockets. Note that the outer $d_{xy}$ hole pocket centered around $\tilde{\Gamma}$ doesn't show much shrinkage, as is also evident from its $\Sigma'(\mathbf{k},\omega_0)$ value which is grazing the level of the new chemical potential. It is worth noting that the self-energy components displayed in Fig. \ref{compare_plots} for TPSC are larger than the magnitudes shown for RPA and FLEX. This is due to the fact that TPSC is able to access higher values of the Coulomb interaction $U/J$ parameters comparable to those obtained from realistic constrained RPA calculations~\cite{ZantoutPRL2019}. In contrast, the interaction parameters used in the weak-coupling RPA/FLEX approach are restricted to smaller magnitudes away from the antiferromagnetic instability point. Despite the large quantitative difference between the self-energy values obtained via TPSC and RPA, we still have qualitatively similar Fermi level features. This is due to the fact that some of the self-energy shifts are undone by the shift of the new chemical potential, which is also much larger in TPSC than in RPA. In Appendix \ref{Zvalues}, we have presented a detailed comparison of quasiparticle weights obtained via RPA and TPSC.

\subsection{Scattering lifetime of quasiparticles} \label{scattering}

\begin{figure*}[hbt!]
    \centering
    \subfloat[$U$ = 0.85 $U_\mathrm{crit}$]{\includegraphics[width=0.49\textwidth]{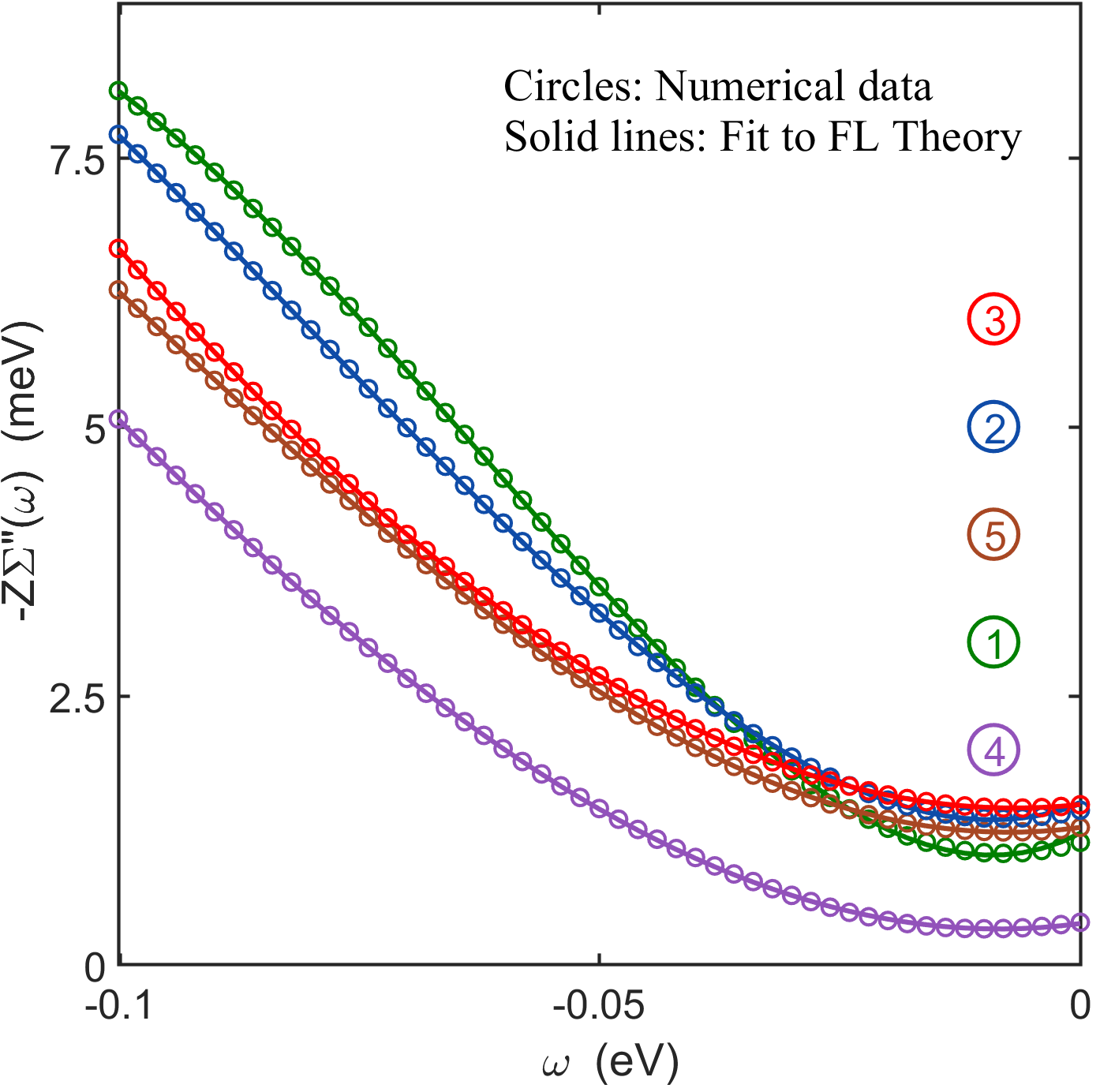}\label{scatter}} \;
    \subfloat[$U$ = 0.95 $U_\mathrm{crit}$]{\includegraphics[width=0.485\textwidth]{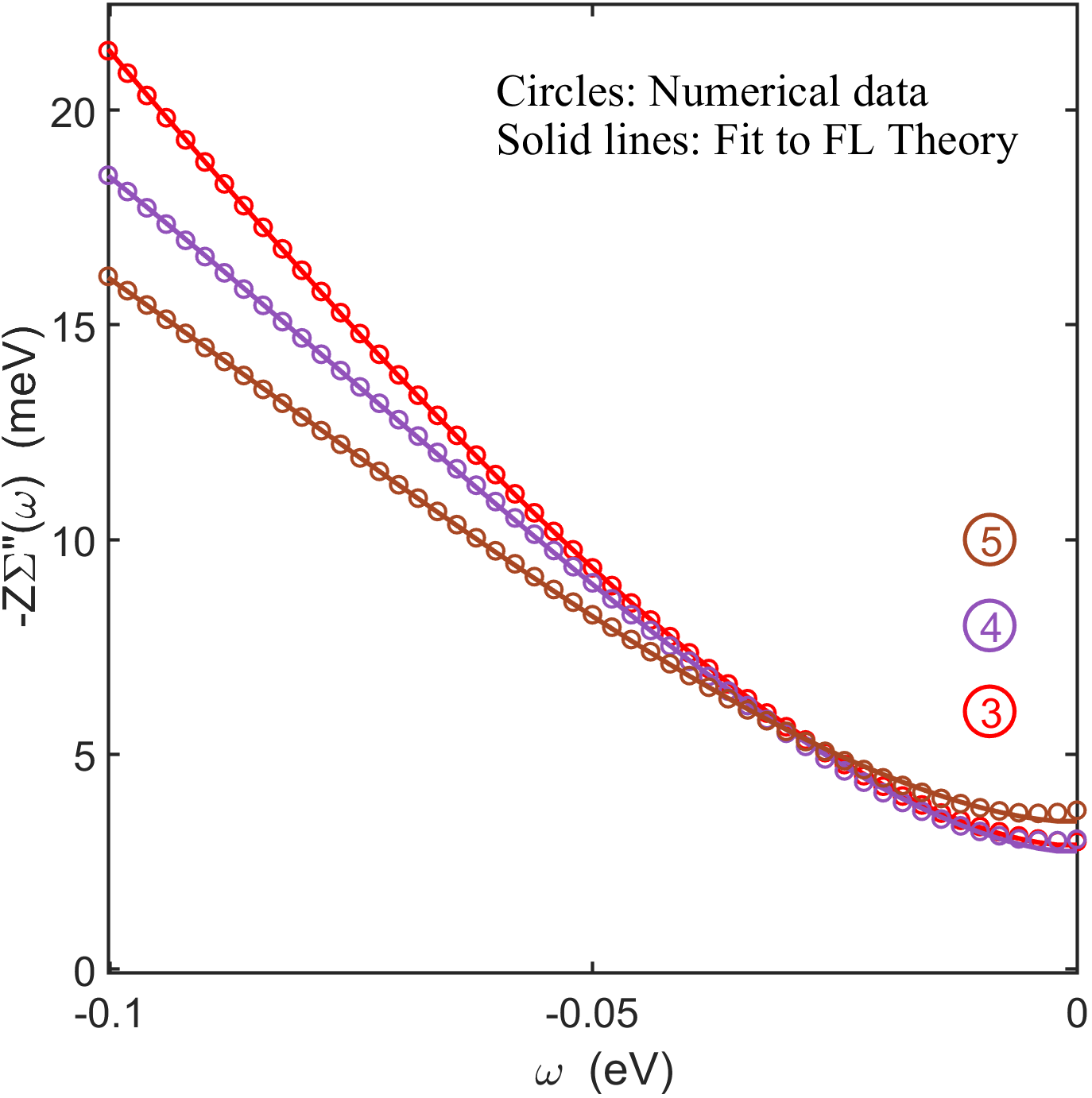}\label{Ucrit_scatter}} 
    \caption{Scattering lifetime of quasiparticles as a function of binding energy $\omega$ evaluated within RPA for (a) $U=0.85U_\mathrm{crit}$, (b) $U=0.95U_\mathrm{crit}$, $J=U/8$ and $T=100$K, at momentum points $1-5$ as denoted in Fig.~\ref{111_2Fe} for LiFeAs. Solid lines are fit to Fermi liquid behavior. Colors refer to particular momentum points as marked in the plots, and are labelled in descending order of their $-Z\Sigma''(0)$ value  from top to bottom.}
    \label{Scatter_rate}
\end{figure*}

In this section, we will address ARPES findings for momentum-dependent scattering lifetime of quasiparticles in LiFeAs for momentum points $1-5$ along $\Gamma-M$ direction as marked in Fig.~\subref*{111_2Fe}. This particular high-symmetry direction includes points on the Fermi surface with orbital weight arising purely from a single specific orbital. This makes it easier to link the behavior at any specific Fermi momentum point to its corresponding orbital. Within Fermi liquid theory, the imaginary part of the retarded self-energy in 2D obeys the Fermi liquid relation within logarithmic accuracy with respect to the binding energy $\omega$ near the Fermi level at temperature $T$~\cite{Maslov2003}:
\begin{align}  \label{marginalFL}
   -\Sigma''(\mathbf{k_F},\omega) \propto  \omega^2 \textnormal{log}|\omega/\Lambda| + \pi^2 T^2 
\end{align}
where $\Lambda \sim E_F$ is some upper energy cutoff. The scattering lifetime of quasiparticles nearby the Fermi level is given by $ -Z(\mathbf{k_F}) \Sigma''(\mathbf{k_F},\omega) $ as discussed earlier in Section \ref{model}. Previous TPSC calculations~\cite{ZantoutPRL2019} have found Fermi-liquid and non-Fermi liquid behavior of quasiparticles.

In ARPES measurements, the quasiparticle lifetimes are typically compared to the dominant quadratic-$\omega$ Fermi liquid term in Eq.\ref{marginalFL} as 
\begin{align}  
   \hbar/\tau = \gamma \left[ (\hbar \omega)^2 + (\pi T)^2 \right]
\end{align}
In Ref.~\onlinecite{Brouet_PRB_2016}, Brouet \textit{et al.} claimed that the Fermi liquid behavior of scattering lifetime depends on the orbital character rather than its position on the Fermi surface (hole or electron pocket). They found that momentum points [3 and 4] dominated by $d_{xy}$ orbital on both hole and electron pockets show strong Fermi liquid behavior with a large $\gamma$, unlike small $\gamma$ value for $d_{xz/yz}$ orbital [points 2 and 5]. Brouet {\it et al.} explained that for a simple metal, the renormalized bandwidth $ZW$ would set a low-energy scale for the coherent part of the spectrum $E_\mathrm{coh}$ to observe Fermi liquid behavior, i.e., $E_\mathrm{coh} \ll ZW$ and $\gamma \sim 1/ZW$~\cite{Georges1996}. In case of $ZW$ being smaller for $d_{xz/yz}$ than for $d_{xy}$, it may not be possible to establish its Fermi liquid regime within current experimental precision. In an independent ARPES study Ref.~\onlinecite{Fink2019}, Fink \textit{et al.} claimed that the scattering lifetime shows strong momentum dependence which can be attributed to nesting conditions and availability of phase-space for interband scattering. They found linear-$\omega$ behavior for $d_{yz/xy}$ orbitals at momentum points 1, 3 (hole) and 5 (electron) while quadratic behavior for $d_{xz/xy}$ orbitals at points 2 (hole) and 4 (electron).  The discrepancy between these two experimental works remains an open question.  

We performed two sets of RPA calculations for Coulomb interaction parameters $U = 0.85U_\mathrm{crit}$ and $U=0.95U_\mathrm{crit}$, with $J=U/8$ and $U_\mathrm{crit}=1.19$ eV being the critical value causing antiferromagnetic instability at $T=100$K. With the larger $U=0.95U_\mathrm{crit}$, the shrinkage of FS pushes the inner hole pockets by almost 15 meV below the Fermi level. As mentioned earlier in Section \ref{bs_results}, since the observed spin-orbit splitting at $\Gamma$ in LiFeAs is of the same order of magnitude, we expect that inclusion of spin-orbit coupling will give rise to a single band  grazing the Fermi level, as seen in ARPES data.

In Fig.\ref{Scatter_rate}, we show the scattering lifetime as a function of frequency for the above two values of $U$. In panel b, we do not show data for momentum points  1 and 2 since in our calculation without spin-orbit coupling, both the inner-hole pockets are pushed below the Fermi level. We find that all our numerical data obtained at different momentum points can be fit to the Fermi liquid behavior as in Eq.\ref{marginalFL}, with fitting coefficients $a,\Lambda,b$: $ -Z \Sigma''(\omega)= a \omega^2 \textnormal{log} |\omega/\Lambda| + b $.


We find that the $d_{xy}$ orbitals become more correlated than $d_{xz/yz}$ as  $U$ is increased, i.e. $Z(d_{xy})$ decreases by a factor of 2.7 compared to 1.7 for $Z(d_{yz})$. The relative magnitudes for $-Z\Sigma''(0)$ changes upon increasing $U$, specifically the $d_{yz}$ electron pocket (point 5) acquires a larger $-Z\Sigma''(0)$ than $d_{xy}$ hole and electron pockets (points 3 and 4). This hierarchy of the $-Z\Sigma''(0)$ values agree with the ARPES findings in Ref.~\onlinecite{Brouet_PRB_2016}. Also, for larger $U$ we find that the fitting coefficient $a$ for Fermi-liquid behavior is larger for momentum points 3 and 4 ($d_{xy}$ orbital content) than for 5 ($d_{yz}$ content), which is also consistent with Ref.~\onlinecite{Brouet_PRB_2016} claims.

Independent of the values of $U$, the  analysis of the RPA scattering lifetime yields Fermi liquid behavior compatible with some experimental data\cite{Brouet_PRB_2016}. The analysis close to criticality also shows  enhancement of orbital differentiation in qualitative agreement with this data. 

\section{Summary and Conclusions} \label{conclusions}

We have performed 2D calculations of the momentum-dependent dynamic self-energy and its corresponding electronic structure renormalization effects for the three prototypical FeSC: LaFeAsO, LiFeAs, and FeSe. We showed that repulsive interband finite-energy scattering processes that are sensitively dependent on the upper/lower edge of the band structure are important in determining the correct sign of the self-energy that causes Fermi surface shrinkage. Treating itinerant spin-fluctuations that result from local Coulomb interactions within both RPA, TPSC and FLEX schemes, our results point towards a universal trend across the families of FeSC, i.e., Fermi surface shrinkage of both hole and electron pockets arising due to non-local, orbitally selective self-energy renormalizations. Although the resulting renormalized Fermi surfaces for Pn-systems LaFeAsO and LiFeAs
agree well with ARPES findings, the Ch-system FeSe does not yield the drastic Fermi surface shrinkage observed in experiments. 

We proposed that nearest-neighbor Coulomb interaction might be
playing a significant role in the band structure renormalization of the
Ch-system. In this context, our calculations show desirable results for FeSe.
Next, we evaluated momentum-dependent scattering lifetime of quasiparticles in
LiFeAs showing Fermi liquid behavior. We have also included detailed discussions about ARPES findings for the
same. We conclude that the inclusion of non-local, orbitally resolved
self-energy effects within the framework of existing spin-fluctuation theories
is an important ingredient to understand the observed electronic structure of
the moderately correlated Pn-systems. Whereas the more strongly correlated
Ch-systems might be dictated by other underlying physical mechanisms, for
example, non-local Coulomb interactions. It is desirable to have further investigation devoted to a
detailed study of the band renormalization in bulk FeSe.

\section*{Acknowledgments}

We acknowledge useful discussions with  S. Backes, L. Benfatto, A. E. B\"ohmer,
V. Borisov, T. Chen, A. Chubukov, D.L. Maslov. S. B. acknowledges support in part through an
appointment at the Goethe-Universit\"at Frankfurt as a short-term research
visiting scientist, sponsored by the Deutscher Akademischer Austauschdienst
(DAAD) and the University of Florida CLAS Dissertation Fellowship funded
by the Charles Vincent and Heidi Cole McLaughlin Endowment. K. B and D. S. and B. M. A. acknowledge support from the Carlsberg
Foundation. B. M. A. acknowledges support from the Independent Research Fund
Denmark grant number DFF-6108-00096. L. F. has received funding from the European Union’s Horizon 2020 research and innovation programme under the Marie Sk\l{}odowska-Curie grant SuperCoop (Grant No 838526).
R. V. acknowledges support by the Deutsche Forschungsgemeinschaft (DFG) through
grant VA117/15-1. P. J. H.  was supported by the Department of Energy under Grant No. DE-FG02-05ER46236. 

\appendix
\section{Numerical details} \label{numeric_details}

Here, we will outline few generic relations that are practical, saves memory and speeds up the actual numerical calculations. For a paramagnetic normal state that is time-reversal invariant, the Green's function obeys the relation:
\begin{align}  
   G_{ps}(\mathbf{k},\omega_{m})^* = G_{sp}(\mathbf{k},-\omega_{m})
\end{align}
Following this, the non-interacting susceptibility obeys the relation $ \chi^0_{pqst}(\mathbf{q},\Omega_{m})^* = \chi^0_{stpq}(\mathbf{q},-\Omega_{m})$. The interaction matrix elements as mentioned in Eq.~\ref{U_matrix} are symmetric under the interchange $U_{pqst} = U_{stpq}$, which allows for the spin and charge RPA susceptibilities and the effective interaction to obey the same symmetry relation like the non-interacting susceptibility:
\begin{align}  
  V_{pqst}(\mathbf{q},\Omega_{m})^* = V_{stpq}(\mathbf{q},-\Omega_{m})
\end{align}
These relations help in restricting our calculations to only one-half of the Matsubara plane while the other half can be symmetry related. With the temperature set to $T=100$K, we used $400$ Matsubara frequencies in the upper half plane for all our calculations. We performed 2D calculations with a $\mathbf{k}$-mesh of $40 \times 40$ in the unfolded $1$-Fe BZ. 

The orbitally resolved non-interacting susceptibility is evaluated using the Lindhard expression:
    \begin{align} 
    \begin{split} 
    & \chi^0_{pqst}(\mathbf{q},\Omega_m) \\
    &= -\frac{1}{N_k \beta}  \sum_{\mathbf{k}\omega_m} G^0_{tq}(\mathbf{k},\omega_m) G^0_{ps}(\mathbf{k+q},\omega_m+\Omega_m) \\	
    & = -\frac{1}{N_k}  \sum_{\mathbf{k} \mu \nu} \frac{ a^t_\mu(\mathbf{k}) a^{q*}_\mu(\mathbf{k}) a^p_\nu(\mathbf{k+q}) a^{s*}_\nu(\mathbf{k+q})}{i\Omega_m + E_\mu(\mathbf{k}) - E_\nu(\mathbf{k+q})} \\
    & \qquad \qquad \qquad \times [f(E_\mu(\mathbf{k})) - f( E_\nu(\mathbf{k+q})) ]
    \end{split}
    \end{align}
where $f(\epsilon) = 1/(e^{\beta \epsilon} + 1)$ is the Fermi-Dirac distribution function. To evaluate the self-energy as in Eq.~\ref{selfenergy}, we used circular convolution theorem with Fast Fourier transform along the momentum space. 

To ensure particle number conservation, one has to evaluate a new chemical potential $\mu$ such that the electron density is a given $n = 2 \displaystyle{\sum_{\nu \mathbf{k}}} f(E_{\nu }(\mathbf{k}))$. Here, $E_{\nu }(\mathbf{k}) $ is the eigenvalue of the unperturbed Hamiltonian $H_0(\mathbf{k})$. With the interacting Green's function, we can evaluate $n$ with the following equation:
\begin{align}  
     n &= 2 \displaystyle{\sum_{\nu \mathbf{k}}} f(E_{\nu }(\mathbf{k})) \notag\\
     &+ \frac{2}{\beta N_k} \displaystyle{\sum_{p,\mathbf{k},\omega_{m}}} \left[ G_{pp}(\mathbf{k},\omega_{m}) - G_{pp}^0(\mathbf{k},\omega_{m}) \right]
\end{align}
In order to produce the plots in Fig.~\ref{main_figs} with high resolution, we have interpolated our Matsubara self-energy data at each frequency point to a $\mathbf{k}$-mesh of $250 \times 250$ points, followed by the evaluation of the corresponding renormalized Green's function. We folded our spectral function from 1-Fe to 2-Fe BZ to present our results in a manner that is easily comparable to experimental data.

\section{RPA calculation on the toy two-band model} \label{RPA_num}

In Section \ref{toy_model}, we illustrated the pocket shrinking mechanism proposed by Ortenzi {\em et al.} in Ref.~\onlinecite{Ortenzi2008} within a toy two-band model for FeSC with two-dimensional (2D) parabolic bands. 
Here, we analyze the case of a momentum-dependent spin-fluctuation interaction and verify that, while it seems natural to associate scattering process at $(\pm \pi,0)$ with interband interactions, in order to recover the shrinking of the pockets one has to consider the region in  momentum space large enough to accommodate scattering processes that connect sufficiently high energy states in an electron (hole) band to those near the top (bottom) of the hole (electron) band.

The hole and electron band dispersions are:
\begin{align}   \label{toy_dispersion}  
     E_\Gamma(\mathbf{k}) &=  - \gamma (k_x^2 + k^2_y ) + \mu \\
     E_{\tilde{\textnormal{M}}}(\mathbf{k}) &= \gamma \left[ \left( k_x-\pi \right) ^2 +  k_y^2 \right] - \mu 
\end{align}
With $\gamma=1.5$, and $\mu=1$, we fix the units of energy in terms of $\mu$ for our current analysis. We ignore the presence of the M-centered electron pocket since it doesn't add any further insight to our analysis. The non-interacting susceptibility in band-basis $(\alpha,\beta)$ is:
\begin{align} \label{bandsus}
    & \chi^0_{\alpha \beta}(\mathbf{q},\Omega_m)\notag \\
    &= -\frac{T}{ N_k }  \sum_{k,\omega_m} G^0_{\alpha}(\mathbf{k},\omega_m) G^0_{\beta}(\mathbf{k+q},\omega_m+\Omega_m) 	
\end{align}
and the charge- and spin-fluctuation parts of the RPA susceptibility is $\chi^{C/S}(\mathbf{q},\Omega_m) =  [ 1 \pm \chi^0(\mathbf{q},\Omega_m) U]^{-1} \chi^0(\mathbf{q},\Omega_m)$. The corresponding particle-hole interaction is:
 \begin{align}
 \begin{split} \label{bandV}
     V_{\alpha \beta}(\mathbf{q},\Omega_{m}) &= \left[ \frac{3}{2} U^2 \chi^S(\mathbf{q},\Omega_{m}) + \frac{1}{2} U^2 \chi^C(\mathbf{q},\Omega_{m}) \right. \\
     &- \left. U^2 \chi^0(\mathbf{q},\Omega_{m}) \right]_{\alpha \beta}
 \end{split}
 \end{align}
A bare Coulomb interaction parameter of $U=8\mu$ and $T= 0.05\mu$ was used for the numerical evaluation. 

\begin{figure}[tb]
    \centering
    \includegraphics[width=\linewidth]{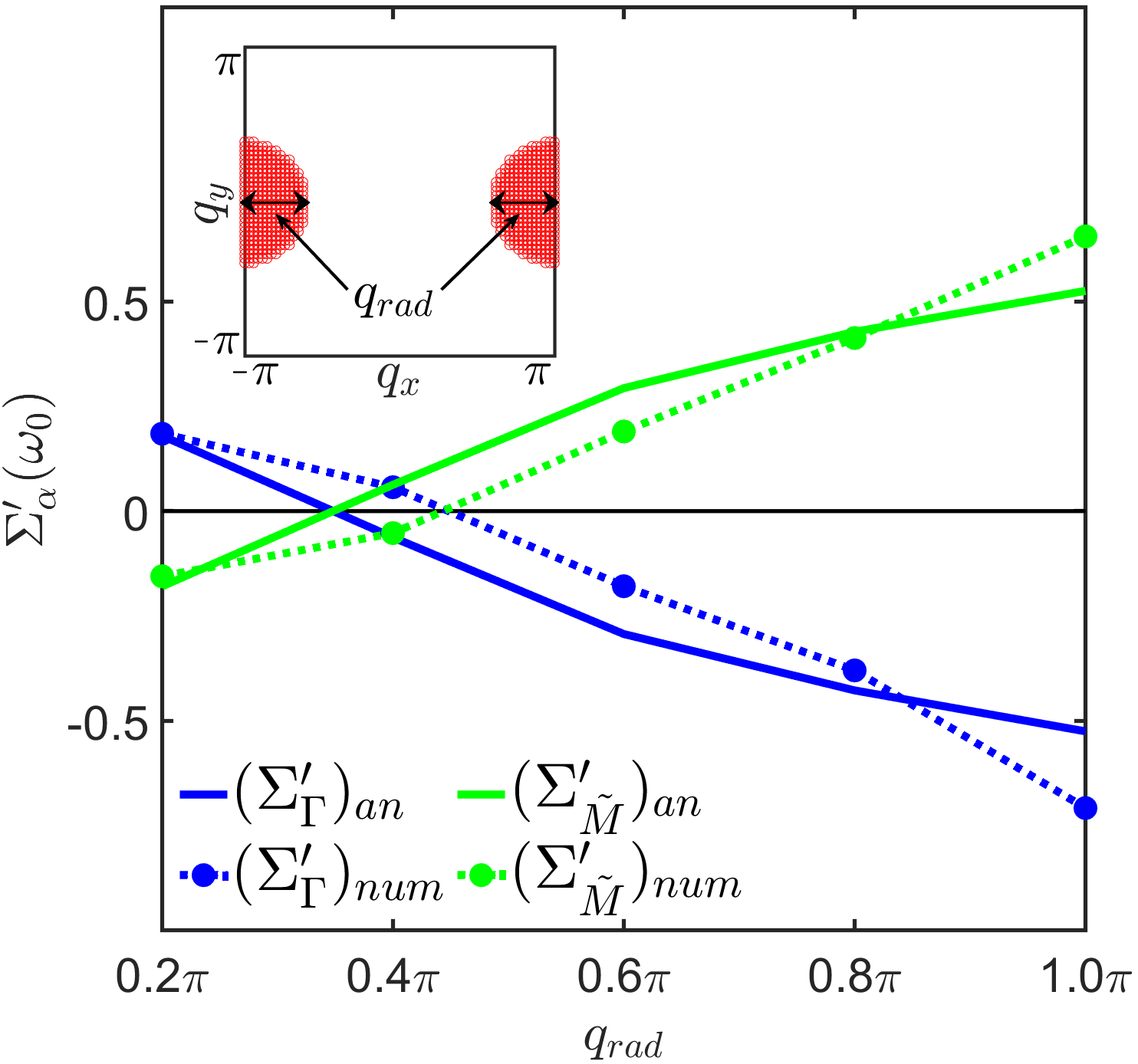}
    \caption{ Plot of the real part of the static self-energy $\Sigma'_\alpha({\omega_0})$ (in units of energy) at $\Gamma$ and $\tilde{\textnormal{M}}$ point as a function of varying $q_\mathrm{rad}$. Both numerical results and analytical predictions have been included. Inset: The shaded region in red depicts the area of \textbf{q}-integration as in Eq.~\ref{SE_appendix} within the radius $ q_\mathrm{rad}$, centered around the $(\pm \pi,0)$ antiferromagnetic wavevector.}
    \label{Sigma_w0}
\end{figure}

We provide both analytical predictions using Eq.~\ref{SE_analytic2}-\ref{SE_analytic1} and numerical results performed using $V_{\alpha \beta}(\mathbf{q},\Omega_{m})$ from Eq.~\ref{bandV} in Eq.~\ref{SE_appendix} for the static self-energy, as a function of $ q_\mathrm{rad}$. In Fig.~\ref{Sigma_w0}, we show that both the analytical $(\Sigma'_\alpha)_\mathrm{an}$ and numerical $(\Sigma'_\alpha)_\mathrm{num}$ results agree qualitatively, changing sign at a specific $ q_\mathrm{rad}$. It eventually yields a desirable value for FS shrinkage, in the limit where $q_\mathrm{rad}$ encompasses the full BZ integration. While the zero of the analytical functions is found exactly at the particle-hole symmetry point (see Eq.~\ref{SE_analytic2}-\ref{SE_analytic1}), the change of sign for the numerical self-energies occurs at a slightly different momenta due to the effect of the momentum dependence of $V_{\alpha  \beta}(\mathbf{q},\Omega_{m})$. The agreement between analytical predictions and numerical results is a strong verification of our correct understanding of the interplay between momentum transfer and finite-energy scattering, and its role in FS shrinkage. With our findings, we re-emphasize that finite energy scattering processes that are sensitively dependent on the upper/lower edge of the band structure are also important in determining the correct sign of the self-energy that causes FS shrinkage.

\section{Comparison of quasiparticle weights obtained via RPA and TPSC} \label{Zvalues}

In this section, we will compare quasiparticle weights $Z(k)$ obtained via different methods and analyze its evolution with respect to the Hubbard interaction. The quasiparticle weight, strictly speaking, is a quantity defined only at the Fermi level. This means only $Z_{d_{xz/yz}}$ and $Z_{d_{xy}}$ can be analyzed for the orbitals carrying weight on the Fermi surface of FeSCs, leaving out $Z_{d_{x^2-y^2}}$ and $Z_{d_{3z^2-r^2}}$.  

The comparison of $Z$ values obtained from RPA and TPSC is not straightforward because the parameters corresponding to $U$, $J$, etc. have slightly different meanings and must be interpreted as renormalized quantities.  For example,  it is well-known that the RPA produces a rather good approximation to the structure of the exact (Quantum Monte Carlo) magnetic susceptibility of the Hubbard model in momentum and frequency space, but requires a strong downward renormalization of  $U$ to do so\cite{Bulut1993}. 
RPA and TPSC approximations therefore work at different scales of $U$, rendering qualitatively similar but quantitatively different values of various quantities. To be more consistent with comparing results,
at the same scale of $U$, 
the TPSC evaluation presented in this section has been obtained using Hubbard-Kanamori form of the interaction matrices as used in RPA with $J$ set to $U/8$, unlike the main text where cRPA values of the interaction parameters were used to compare with Ref. \onlinecite{ZantoutPRL2019}.
  
\begin{figure*}[hbt!]
    \centering
    \subfloat[]{\includegraphics[width=0.31\textwidth]{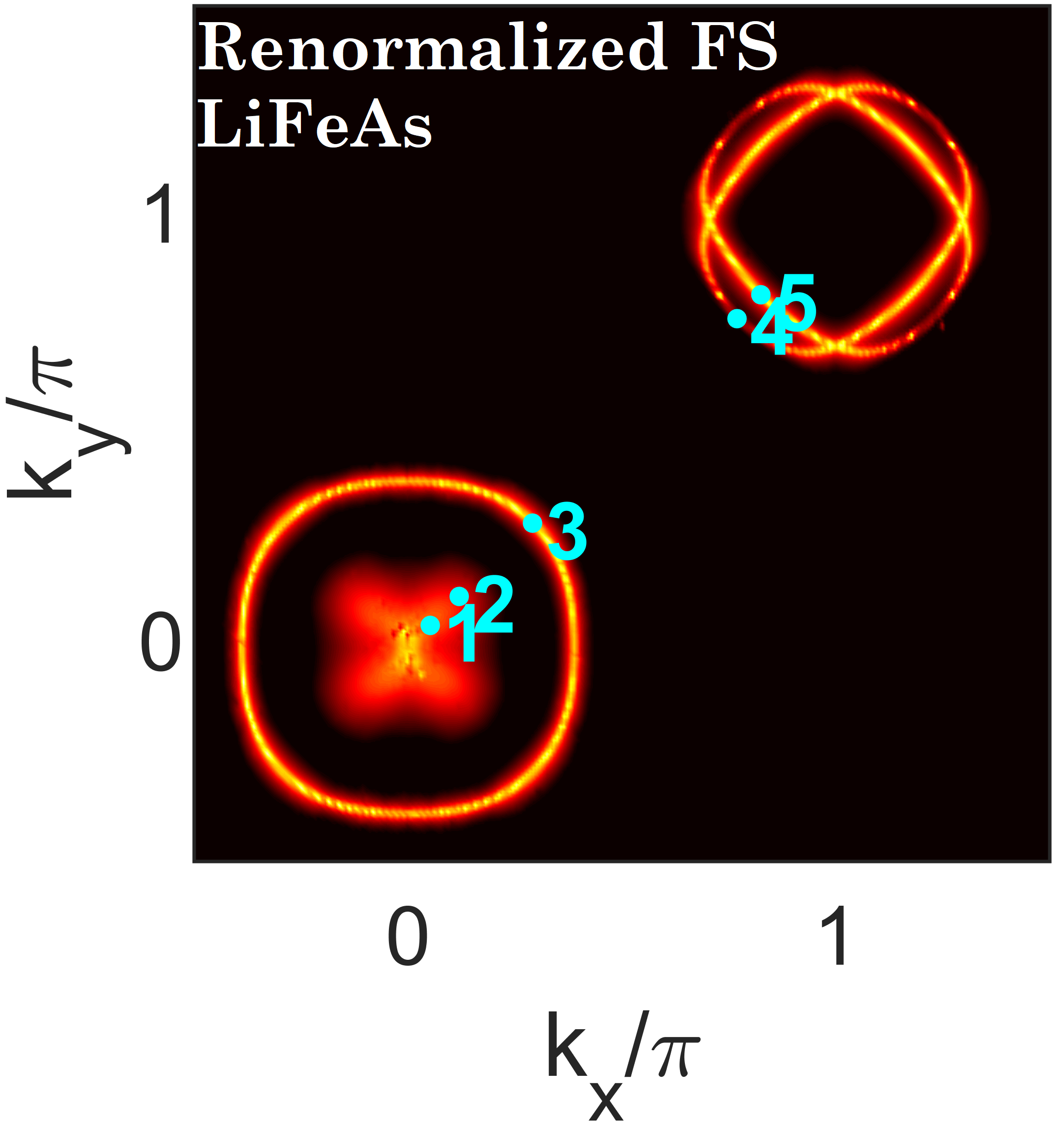}\label{LiFeAs_FS}} \;
    \subfloat[]{\includegraphics[width=0.33\textwidth]{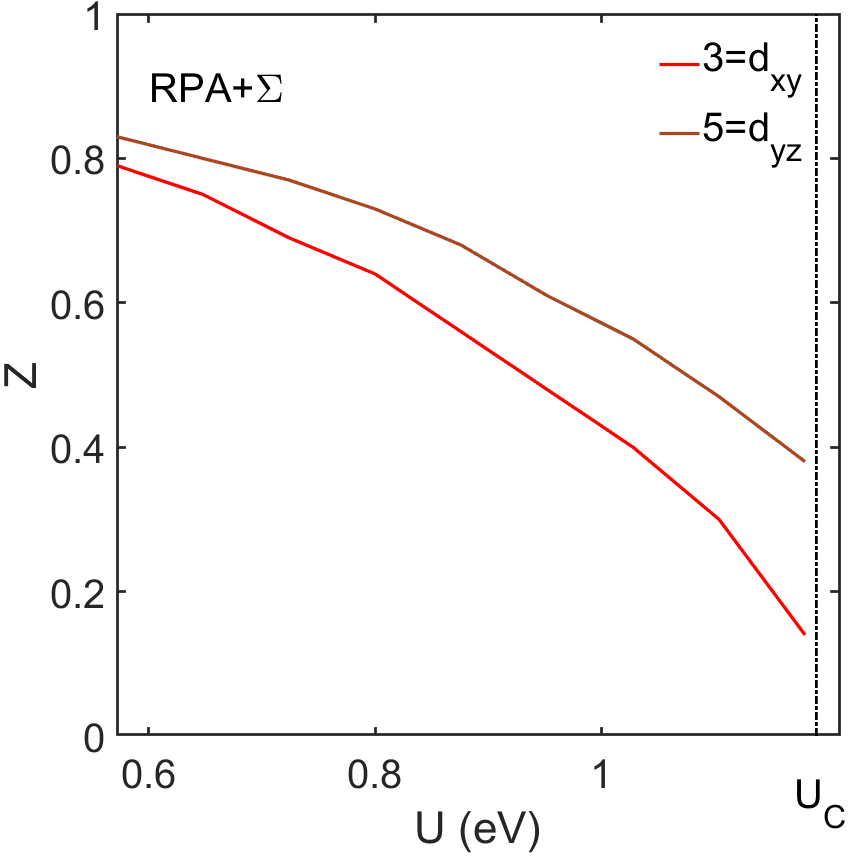}\label{LiFeAs_ZRPA}} 
    \subfloat[]{\includegraphics[width=0.335\textwidth]{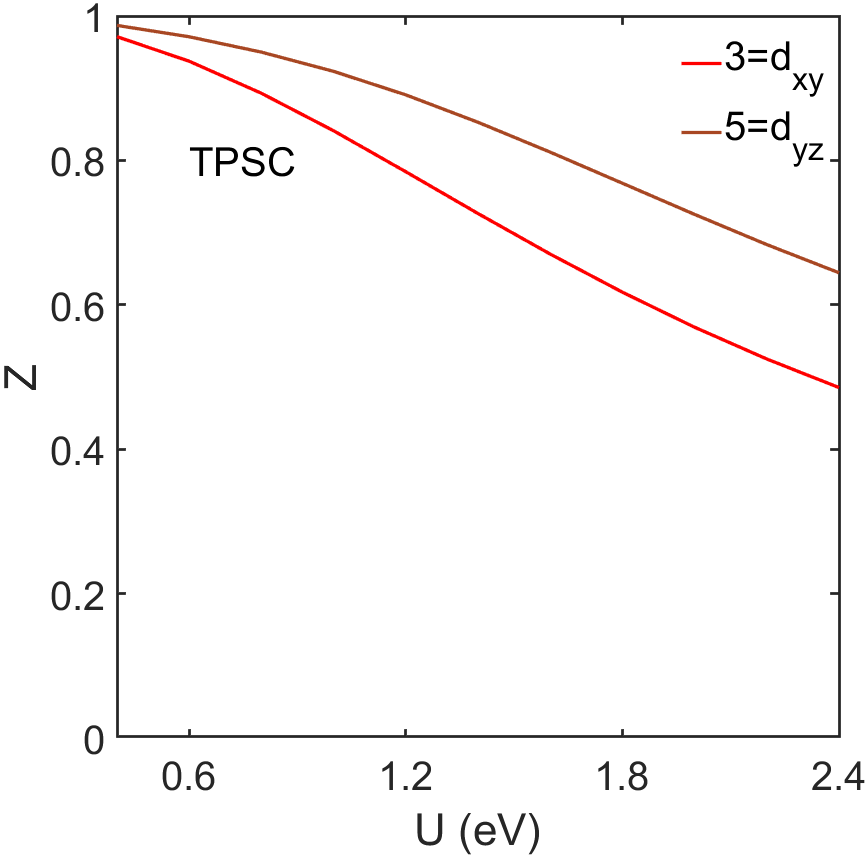}\label{LiFeAs_ZTPSC}}
    \caption{(a) Renormalized Fermi surface of LiFeAs evaluated via RPA at $U=0.8$ eV and $J=U/8$. (b) Plot of the evolution of quasiparticle weight $Z$ with increasing $U$ value evaluated via RPA for momentum points 3 and 5 as marked in figure (a). (c) Same as in (b) but for TPSC calculation using Hubbard-Kanamori interaction matrices.}
    \label{Z_vs_U}
\end{figure*}

In Fig.~\subref*{LiFeAs_ZRPA} and ~\subref*{LiFeAs_ZTPSC}, we have shown the evolution of $Z$ as a function of $U$ for LiFeAs from RPA and TPSC calculations, respectively. As already discussed in Section \ref{scattering}, upon increasing $U$ value, the two inner hole pockets of LiFeAs are pushed below the Fermi level making momentum points 1 and 2 (see Fig.~\subref*{LiFeAs_FS}) vanish. Hence, data for these points are not shown.   The $U_C = 1.19$ eV marked in Fig.~\subref*{LiFeAs_ZRPA} signifies the critical $U$ value causing antiferromagnetic instability in RPA calculation. On the other hand, TPSC calculations are not inhibited by the same $U_C$ at such low values of $U$.  One can see the expected trend of $Z$ decreasing as $U$ increases. For all the cases, we see that $Z_{d_{xy}} < Z_{d_{xz/yz}}$, but the TPSC $Z$ values are much larger than the RPA $Z$ values. Point 4 has $d_{xy}$ orbital content which is the same as point 3, hence not repeated in the plot.

\bibliography{references_orb_sel}{}

\end{document}